\documentclass{aa}

\usepackage{graphicx}
\usepackage{natbib}
\usepackage{float}
\usepackage{color}
\usepackage{multirow}
\usepackage[para,online,flushleft]{threeparttable}

\usepackage{txfonts}
\usepackage[]{hyperref}
\hypersetup{
  colorlinks,
  citecolor=blue,
  linkcolor=red,
  urlcolor=blue}
\usepackage[dvipsnames]{xcolor}

\bibpunct{(}{)}{;}{a}{}{,} 
\newcommand{\psr}{PSR~J0955$-$6150}

\begin{document} 

   \title{The eccentric millisecond pulsar, \psr}
   \subtitle{I: Pulse profile analysis, mass measurements and constraints on binary evolution}
   \titlerunning{Mass measurements of \psr}
   \author{M.~Serylak \inst{\ref{skao}, \ref{sarao}, \ref{uwc}}
           \and V.~Venkatraman~Krishnan \inst{\ref{mpifr}}
           \and P.~C.~C.~Freire \inst{\ref{mpifr}}
           \and T.~M.~Tauris \inst{\ref{aalborg}, \ref{mpifr}}
           \and M.~Kramer \inst{\ref{mpifr}}
           \and M.~Geyer \inst{\ref{sarao}}
           \and A.~Parthasarathy \inst{\ref{mpifr}}
           \and M.~Bailes \inst{\ref{swin},\ref{ozgrav}}
           \and M.~C.~i~Bernadich \inst{\ref{mpifr}}
           \and S.~Buchner \inst{\ref{sarao}}
           \and M.~Burgay \inst{\ref{inaf}}
           \and F.~Camilo \inst{\ref{sarao}}
           \and A.~Karastergiou \inst{\ref{oxford}, \ref{rhodes}}
           \and M.~E.~Lower \inst{\ref{csiro}, \ref{swin}}
           \and A.~Possenti \inst{\ref{inaf}}
           \and D.~J. Reardon \inst{\ref{swin},\ref{ozgrav}}
           \and R.~M. Shannon \inst{\ref{swin},\ref{ozgrav}}
           \and R.~Spiewak \inst{\ref{jbca},\ref{swin},\ref{ozgrav}} 
           \and I.~H.~Stairs \inst{\ref{ubc}}
           \and W.~van~Straten \inst{\ref{aut}}
          }

   \institute{
              SKA Observatory, Jodrell Bank, Lower Withington, Macclesfield, SK11 9FT, United Kingdom\label{skao}
              \and
              South African Radio Astronomy Observatory, 2 Fir Street, Black River Park, Observatory 7925, South Africa\label{sarao}
              \and
              Department of Physics and Astronomy, University of the Western Cape, Bellville, Cape Town, 7535, South Africa\label{uwc}
              \and
              Max-Planck-Institut f\"{u}r Radioastronomie, Auf dem H\"{u}gel 69, D-53121 Bonn, Germany\label{mpifr}
              \and 
               Department of Materials and Production, Aalborg University, Skjernvej 4A, DK-9220 Aalborg~{\O}st, Denmark\label{aalborg}
              \and
              Centre for Astrophysics and Supercomputing, Swinburne University of Technology, PO Box 218, Hawthorn, VIC 3122, Australia\label{swin} \and
              ARC Centre of Excellence for Gravitational Wave Discovery (OzGrav) \label{ozgrav}
              \and
              INAF-Osservatorio Astronomico di Cagliari, via della Scienza 5, 09047 Selargius, Italy \label{inaf}
              \and Department of Astrophysics, University of Oxford, Denys Wilkinson building, Keble Road, Oxford OX1 3RH, UK \label{oxford}
              \and
              Department of Physics and Electronics, Rhodes University, PO Box 94, Grahamstown 6140, South Africa \label{rhodes}
              \and
              CSIRO, Space and Astronomy, PO Box 76, Epping, NSW 1710, Australia\label{csiro}
              \and              
              Jodrell Bank Centre for Astrophysics, Department of Physics and Astronomy, University of Manchester, Manchester M13 9PL, UK\label{jbca} 
              \and
              Dept. of Physics and Astronomy, University of British Columbia, 6224 Agricultural Road, Vancouver, BC, V6T 1Z1 Canada\label{ubc} 
              \and
              Institute for Radio Astronomy \& Space Research, Auckland University of Technology, Private Bag 92006, Auckland 1142, New Zealand\label{aut}\\
              \email{maciej.serylak@skao.int; vkrishnan@mpifr-bonn.mpg.de}
            }

   \date{Received --; accepted --}

\abstract{}{}{}{}{} 
\abstract
   {\psr\, is a member of an enigmatic class of eccentric MSP+He~WD systems (eMSPs), whose binary evolution is poorly understood and believed to be strikingly different to that of traditional MSP+He~WD systems in circular orbits.}
   {Measuring the masses of the stars in this system is important for testing the different hypotheses for the formation of eMSPs.}
   {We have carried out timing observations of this pulsar with the Parkes radio telescope using the 20-cm multibeam and ultra-wide bandwidth low-frequency (UWL) receivers, and the L-band receiver of the MeerKAT radio telescope. The pulse profiles were flux and polarisation calibrated, and a Rotating Vector Model (RVM) was fitted to the position angle of the linear polarisation of the combined MeerKAT data. Pulse times of arrival (ToAs) were obtained from these using standard pulsar analysis techniques and analysed using the {\sc tempo2} timing software.}
   {Our observations reveal a strong frequency evolution of this millisecond pulsar's intensity, with a flux density spectral index ($\alpha$) of $-3.13(2)$. The improved sensitivity of MeerKAT has resulted in a $>10$-fold improvement in the timing precision obtained compared to our older Parkes observations. This, combined with the 8-year timing baseline, has allowed precise measurements of a very small proper motion and three orbital ``post-Keplerian'' parameters, namely the rate of advance of periastron, $\dot{\omega} = 0.00152(1) \, \deg \rm yr^{-1}$ and the orthometric Shapiro delay parameters: $h_3 = 0.89(7) \, \mu$s and $\varsigma = 0.88(2)$. Assuming general relativity, we obtain $M_{p} = 1.71(2) \, M_{\odot}$ for the mass of the pulsar and $M_{c} = 0.254(2) \, M_{\odot}$ for the mass of the companion; the orbital inclination is 83.2(4) degrees. Crucially, assuming that the position angle of the linear polarisation follows the rotating vector model, we find that the spin axis has a misalignment relative to the orbital angular momentum of $> 4.8 \deg$ at 99\% CI.}
   {While the value of $M_{\rm p}$ falls well within the wide range observed in eMSPs, $M_{\rm c}$ is significantly smaller than expected by several formation hypotheses proposed, which are therefore unlikely to be correct and can be ruled out. $M_{\rm c}$ is also significantly different from the expected value for an ideal low mass X-ray binary evolution scenario. If the misalignment between the spin axis of the pulsar and the orbital angular momentum is to be believed, it suggests that the unknown process that created the orbital eccentricity of the binary was also capable of changing its orbital orientation, an important evidence for understanding the origin of eMSPs.}
   \maketitle
\section{Introduction}
\label{section:Introduction}

Radio pulsars are unique objects in astronomy: they are neutron stars (NSs) that emit a regular train of radio pulses whose times of arrival at the telescope (TOAs) can be measured with great precision \citep{Lorimer&Kramer2005}. This allows the determination of extremely precise spin periods (sometimes better than a few femtoseconds), the rate of variation of this spin period, and astrometric parameters such as positions and proper motions that are precise to microarcseconds, similar to the precision obtained from very large baseline interferometry (VLBI). In the case of pulsars in binary systems, pulsar timing allows exquisitely precise measurements of their orbital motion, which can be used for precise measurements of the components of the binary \citep{2016ARA&A..54..401O} and tests of gravity theories \citep{Wex2014,2021PhRvX..11d1050K}.

The  MeerKAT 64-dish array \citep{2009IEEEP..97.1522J} now provides excellent sensitivity (and thus timing precision) for Southern radio pulsars. Precision pulsar timing is carried out under the MeerTime large science project (LSP; \citealt{BailesEtAl2020}); the first results are extremely promising and show a performance that is significantly better than expected. Within the MeerTime LSP, there is a program targeting relativistic binary pulsars, henceforth ``RelBin". The objective of this program is to use the excellent timing precision provided by the MeerKAT to improve the measurement or detection of new relativistic effects in the orbital motion of known binary pulsars, with the aim of a) increasing the number of neutron stars (NS) with precise mass measurements and b) increasing the number, nature and precision of pulsar tests of gravity theories (for details, see \citealt{ksv+21}).

One of the early additions to the RelBin program was \psr, a binary millisecond pulsar with a low-mass companion and a 24.6-d orbit with an unusual eccentricity ($e = 0.11$) that was discovered with the CSIRO Parkes 64-m radio-telescope (recently given the indigenous Wiradjuri name ``Murriyang'') in a survey of unassociated {\em Fermi} sources \citep{CamiloEtAl2015}. In that paper, no phase-coherent timing solution was presented for this pulsar, owing to its extreme faintness, only an orbital solution derived from the observed variation of the spin period (the ``Doppler'' method). We were able to derive a phase-connected timing solution for this pulsar based on its Parkes long term timing data. Still, even at that stage, with orbital parameters thousands of times more precise than those derived from the Doppler method, we could only detect one relativistic effect in the orbit (the advance of periastron, which proceeds at a rate known as $\dot{\omega}$), and that with low significance. Because of the limited timing precision, no other relativistic effects on the timing of the pulses could be detected. This unpublished timing solution was the basis for the timing solution presented in this work.

\begin{table*}[h]
 \caption[]{
 \label{table:eccentric_MSPs}
 Parameters for the eccentric MSPs known in the Galactic disk. Note the similarity of the parameters for the first {\bf five} pulsars, and how they differ significantly from PSR~J1903+0327. The first five systems and PSR~J1146$-$6610 all have mass functions, spin periods and period derivatives typical of MSPs with He~WD companions. The 8th column states the He~WD mass interval expected from the $P_{\rm b}-M_{\rm WD}$ correlation of \cite{1999A&A...350..928T}. For PSR~J1903+0327, the mass uncertainties refer to a 99.7\,\% confidence limit; the companion of that pulsar is a main-sequence star. For the other systems, the companions are presumably He~WDs; this has been confirmed in the case of PSR~J2234+0611 by \cite{2016ApJ...830...36A}.}
 \centering
  \begin{tabular}{c r l l l l l l c}
   \hline
   \hline
   PSR & $P$ (ms) & $P_{\rm b}$ (d) & $e$ & $M_{\rm T}$ ($M_{\odot}$) & $M_{\rm p}$ ($M_{\odot}$) & $M_{\rm c}$ ($M_{\odot}$) & $M_{\rm theo}$ ($M_{\odot}$) & References \\
   \hline
   J0955$-$6150 &  1.9993 & 24.5784 & 0.1175 & 1.96(3)                      & 1.71(2)                   & 0.254(2)                     & 0.271--0.300& 1, 2   \\
   J1618$-$3921 & 11.9873 & 22.7456 & 0.0274 & -                            & -                         & -                            & 0.269--0.297& 3, 4   \\
   J1946+3417   &  3.1701 & 27.0199 & 0.1345 & 2.094(22)                    & 1.827(13)                 & 0.2654(13)                   & 0.275--0.303& 5, 6   \\
   J1950+2414   &  4.3048 & 22.1914 & 0.0798 & 1.779(25)                    & 1.496(23)                 & $0.2795^{+0.0046}_{-0.0038}$ & 0.268--0.296& 7, 8   \\
   J2234+0611   &  3.5766 & 32.0014 & 0.1293 & $1.6518^{+0.0033}_{-0.0035}$ & $1.353^{+0.014}_{-0.017}$ & $0.298^{+0.015}_{-0.012}$    & 0.281--0.310& 9, 10  \\
   \hline
   J1146$-$6610 &  3.7223 & 62.7712 & 0.0074 & -                            & -                         & -                            & 0.307--0.339 & 11     \\
   \hline
   J1903+0327   &  2.1499 & 95.1741 & 0.4367 & 2.697(29)                    & 1.667(21)                 & 1.029(8)                     & ---          & 12, 13 \\
   \hline
  \end{tabular}
  \tablebib{(1)~\cite{CamiloEtAl2015}; (2)~this work; (3)~\cite{2018AA...612A..78O}; (4)~\cite{2001ApJ...553..801E}; (5)~\cite{2013MNRAS.435.2234B}; (6)~\cite{2017MNRAS.465.1711B}; (7)~\cite{2015ApJ...806..140K}; (8)~\cite{2019ApJ...881..165Z}; (9)~\cite{2013ApJ...775...51D}; (10)~\cite{2019ApJ...870...74S}; (11)~\cite{2021MNRAS.507.5303L}; (12)~\cite{2008Sci...320.1309C}; (13)~\cite{2011MNRAS.412.2763F}
  }
\end{table*}

This pulsar was added to the RelBin program because of the prospect of a high-precision mass measurement. Indeed, for three systems similar to \psr \, (see Table~\ref{table:eccentric_MSPs}), the large orbital eccentricities and the high precision of Arecibo timing allowed precise measurements of $\dot{\omega}$ and a measurement of a relativistic, orbital-phase dependent delay in the arrival times of the pulses, known as the Shapiro delay \citep{1964PhRvL..13..789S}, which is a direct consequence of the fact that the radio waves from the pulsar propagate in a curvature of space-time. The combination of these effects is enough for a precise determination of the component masses, even at low inclinations \citep{2019ApJ...870...74S}. It was expected that the precise MeerKAT timing of \psr\,  would allow similarly precise measurements of these relativistic effects in this system and therefore yield a precise measurement of its component masses. As described in detail below, the quality of the MeerKAT L-band detections of this pulsar exceeded all expectations, and precise masses, and much else, can be derived from these detections.

The remainder of this paper is organised as follows: in section~\ref{sec:eMSPs}, we will set the stage by elaborating on the nature of eccentric MSPs binaries like \psr, and discuss what was previously known about their evolution. In section~\ref{section:Observations}, we describe the radio observations of this pulsar and how the resultant data were processed. In section~\ref{section:profile_analysis}, we present the results from our analysis of the pulsar's radio emission, with a focus on its pulse profile: its flux, polarisation, spectral index and scattering measurements. In section~\ref{sec:timing} we present our timing results, these include a discussion of the relativistic effects detected in this system and mass estimates and orbital inclination based on these. In section~\ref{sec:rvm}, we discuss some of the constraints the orientation of the spin axis of the pulsar that result from our polarimetric measurements and compare this orientation with the constraints on the orbital geometry, finding a misalignment between the spin axis of the pulsar and the orbital angular momentum. Finally, in section~\ref{section:Discussion}, we use our mass measurements and the aforementioned orbital misalignment to evaluate the different hypotheses proposed for the evolution of the eccentric MSP+He~WD systems. We summarise our results in section~\ref{sec:summary}.

\section{Eccentric millisecond pulsars}
\label{sec:eMSPs}
\subsection{\psr, a peculiar system}

As mentioned above, PSR~J0955$-$6150 was discovered in a Parkes survey of unidentified Fermi-LAT sources \citep{CamiloEtAl2015}, it coincides with LAT source 3FGL J0955.6$-$6148 \citep{2015ApJS..218...23A}. Such surveys are part of a successful global effort to find pulsar counterparts to unidentified Fermi-LAT sources, many of these are gamma-ray MSPs \citep{2012arXiv1205.3089R}. The pulsar has a spin period of 1.99 ms, hence is a recycled ``millisecond pulsar'', henceforth "MSP" (this is confirmed by the small value for the spin-down, to be discussed later). Like most MSPs, it is in a binary system, in this case with an orbital period $P_{\rm b} \sim 24.58\,\rm d$; the companion has a relatively low mass and is presumably a helium white dwarf star (He~WD). The unusual feature of this system is its orbital eccentricity, $e \simeq 0.12$, which is much larger than the eccentricities of most MSP+He~WD systems. However, these properties are very similar to those of a few systems discovered over the last decade (listed in Table~\ref{table:eccentric_MSPs}). Those systems have orbital periods between 22 and 32 d and orbital eccentricities of the order of 0.1; we will refer to these as eMSP systems.

A recent possible addition to this category, PSR~J1146$-$6610, has an orbital period that is twice as large and an orbital eccentricity that is one order of magnitude smaller than those of the other eMSPs, so it is still unclear to what extent this is related to them \citep{2021MNRAS.507.5303L}. However, its eccentricity is still 2 orders of magnitude larger than that of other MSP - He WD systems with the same orbital period; for this reason we also list it in Table~\ref{table:eccentric_MSPs}.

The evolution of MSP+He~WD systems generally involves a long period ($\sim {\rm Gyr}$) of accretion of matter onto the NS from a low-mass star \citep[][and references therein]{tv22}. During this stage, the system is detectable as a low-mass X-ray binary (LMXB). In these systems, the orbits are invariably circularised by the tidal interaction with the red-giant companion \citep{1992RSPTA.341...39P}. After Roche-lobe overflow (RLO), the pulsar becomes a radio MSP, and the companion becomes a WD. 

In high-mass X-ray binary systems, where the companion star is massive enough to terminate its life in a supernova (SN), the orbit is disturbed by the instantaneous mass loss and the momentum kick imparted onto the newborn NS. In this case, a double NS system is formed if the binary remains bound. Such massive companions evolve much faster, therefore the RLO episode is much shorter. The consequence is that pulsars in these systems do not spin as fast as ``fully'' recycled MSPs like \psr\, (the fastest pulsar in a Galactic disk double NS system, PSR~J1946+2052, has a spin period of $\sim$17 ms \citep{2018ApJ...854L..22S}; see also \citet{tkf+17} for further discussions).

In globular clusters, some of the MSP+He~WD systems acquire eccentric orbits, but these result either from close encounters with other stars in these clusters \citep{1992RSPTA.341...39P} or, in some extreme cases they result from the replacement of the former mass donors with much more massive degenerate companions, possibly NSs (like NGC~1851A, \citealt{2019MNRAS.490.3860R}, NGC~6544B, \citealt{2012ApJ...745..109L}, NGC 6624G, \citealt{2021MNRAS.504.1407R} and NGC~6652A, \citealt{2015ApJ...807L..23D}). Outside globular clusters, the vast majority of all binary millisecond pulsars have very small residual eccentricities consistent with the expectation for the gravitational interaction of the neutron star with the convection cells in the envelope of the WD progenitor star during the last stages of its evolution \citep{1992RSPTA.341...39P}. The exceptions are the systems in Table~\ref{table:eccentric_MSPs}: their orbital eccentricities are 2 - 3 orders of magnitude larger than the prediction.

\subsection{Formation of the eccentric MSPs in the Galaxy}

The presence of eMSPs with their eccentric orbits in the Galactic disk represents a deviation from the predictions of standard evolutionary theory. In the case of PSR~J1903+0327, a 2.15-ms pulsar in an eccentric ($e \, = \, 0.43$) 95-d orbit \citep{2008Sci...320.1309C}, the companion star turned out to be a $1.03 \, M_{\odot}$ main-sequence star \citep{2011MNRAS.412.2763F,2012ApJ...744..183K}. A detailed analysis of the characteristics of this system leads to the conclusion that it very likely originated as a triple system, which later became unstable, a conclusion that was reached by empirical exclusion of all likely alternatives \citep{2011MNRAS.412.2763F} and independently by numerical simulations \citep{2011ApJ...734...55P,2012MNRAS.424.2914P}.

This mechanism is therefore a candidate for the formation of the remaining eMSPs. However, as first pointed out by \cite{2014MNRAS.438L..86F} and \cite{2015ApJ...806..140K}, their characteristics are too similar for such an origin as the disruption of a triple system, which is generally a chaotic process: their observed orbital periods exist within a narrow range between 22 and 32~d, and the orbital eccentricities are also seen in a narrow range between 0.027 and 0.14. It is this  similarity of characteristics that defines, for the moment, the class of eMSPs.

Furthermore, the chaotic destruction of a triple system would, generally, lead to the formation of a binary system consisting of the MSP with the more massive of the remaining two stars as the companion. While this is the case for PSR~J1903+0327, which has an unusually massive (and un-evolved) main sequence companion --- a unique case among the known MSPs in the Galactic disk --- this is not the case for the remaining binaries, where the companion masses measured to date (presented in Table~\ref{table:eccentric_MSPs}) are as expected for He~WD companions for these orbital periods \citep{1999A&A...350..928T}, or slightly smaller (e.g., \citealt{2017MNRAS.465.1711B}). In one case, the relatively nearby PSR~J2234+0611 \citep{2013ApJ...775...51D,2019ApJ...870...74S}, the He~WD nature of the companion has been confirmed by optical observations \citep{2016ApJ...830...36A}.

If not formed in the disruption of a triple system, then how did eMSPs form? The similarity of their orbital periods and eccentricities suggests a process with a relatively fixed outcome. There are at least 5 hypotheses presented in the literature so far. The first two involve a phase transition of the object that becomes the present-day MSP \citep{2014MNRAS.438L..86F,2015ApJ...807...41J}. The next two involve thermonuclear runaway burning \citep{2014ApJ...797L..24A,2021ApJ...909..161H}. Finally, a more recent hypothesis involves resonant convection \citep{Ginzburg21}.

In the first hypothesis by \cite{2014MNRAS.438L..86F}, the phase transition was from a rotationally-delayed accretion-induced collapse (RD-AIC) of a massive WD with a mass above that of the Chandrasekhar limit. The delay of the AIC is possible if the WD is sustained by a fast rotation, where the resulting centrifugal forces prevent the collapse of the WD during RLO. As this rotation slows down after mass accretion has ceased, the centrifugal forces decrease and the collapse of the massive WD becomes inevitable, although whether this can form a MSP directly is still debatable. The second hypothesis involves a phase transition within the neutron star \citep{2015ApJ...807...41J}, for instance, between normal neutron matter and quark matter, also caused by its spin-down and associated decrease in centrifugal forces. Both hypotheses predict interesting optical transient counterparts and produce MSPs within a relatively narrow range of masses: the first hypothesis produces MSPs with masses below $1.32\;M_{\odot}$ (unless differential rotation was at work in the collapsing WD, allowing for significantly more massive NSs to be produced), the second hypothesis would produce MSPs with larger masses, corresponding to the central density at which nuclear matter undergoes a phase transition to a more compact type of matter; but the exact values of the transition mass would depend on the detailed microscopic model for super-dense nuclear matter. Both hypotheses are therefore unable to describe the large range of NS masses already observed for eMSP systems. 

The third and fourth hypothesis, put forward by \cite{2014ApJ...797L..24A} and \cite{2021ApJ...909..161H}, respectively, rely on the expectation that for WDs within the range of the masses predicted for this interval of orbital periods, there should be significant H-shell flashes \citep[i.e. thermonuclear runaway burning episodes,][]{2013A&A...557A..19A,2016A&A...595A..35I} near the surface of the (proto) He~WD. In the hypothesis of \cite{2014ApJ...797L..24A}, this ejects enough material to produce a circumbinary disk that perturbs the orbit and fosters significant eccentricity. In the hypothesis of \cite{2021ApJ...909..161H}, the eccentricity is produced by the ejection of material from the region(s) where runaway nuclear burning is happening --- in effect a ``thermonuclear rocket''. According to this hypothesis, these burning episodes and their associated net ``kicks'' (of order $1-8\;{\rm km\,s}^{-1}$) can then create not only a measurable change in eccentricity, but also a change in the orbital period. As an example, following the recipe of \citet{2021ApJ...909..161H}, applying a NS+WD system with total mass similar to that of \psr\, ($M=M_{\rm p}+(M_{\rm c}+\Delta M)=1.71\;M_\odot + 0.255\;M_\odot = 1.965\;M_\odot$), a relatively large instantaneous kick of $w=8\;{\rm km\,s}^{-1}$, an amount of ejected mass of $\Delta M=10^{-3}\;M_\odot$ and a pre-flash orbital period of $P_{\rm b}=20\;{}\rm d$, results in a post-flash orbital period between $P_{\rm b}\simeq 16.1-26.5\;{\rm d}$, depending on kick direction.

While all hypotheses acknowledge the need for a $P_{\rm b}-M_{\rm WD}$ correlation \citep[such as the one given by][]{1999A&A...350..928T}, none of the helium-flash hypotheses make any specific predictions for the MSP masses; merely that they should reflect the range of masses observed for other MSPs, which seems to be the case \citep{2016ARA&A..54..401O,2016arXiv160501665A}. In the fourth hypothesis, with multiple rocket episodes, we might also have significant changes in the orbital plane, producing a misalignment with the spin axis of the pulsar. This is an important prediction that will be especially important for this work.

Finally, a more recent hypothesis for the formation of the eMSPs has been proposed by \cite{Ginzburg21} who expanded on the work of \citet{1992RSPTA.341...39P} and argue that formation of eMSPs might be due to resonant convection. In the earlier work, it had already been noticed that the timescale of convective eddies within the red giant WD progenitors is about 25~d, which is, again, similar to that of the orbital periods of eMSP systems. To explain their high eccentricities, \cite{Ginzburg21} postulate a coherent resonance between the orbital period and the convective eddies in the red giant progenitors, which drives the anomalously large eccentricities in eMSPs by convective flows.

\section{Observations}
\label{section:Observations}

\begin{figure*}
 \centering
 \includegraphics[scale=0.45]{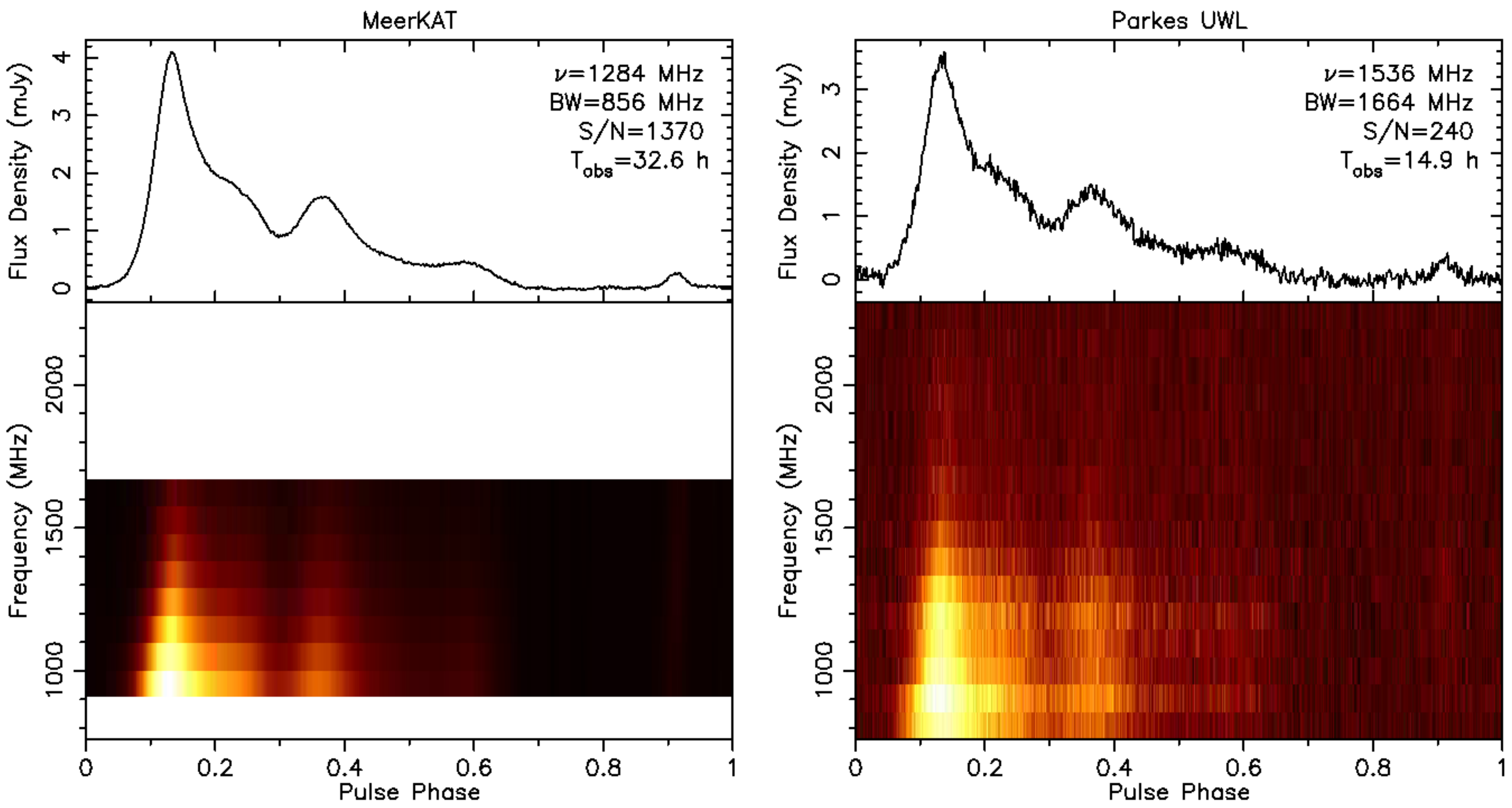}
 \caption{Flux calibrated total intensity profiles made from observations made with MeerKAT (left plot) and Parkes (right plot) radio telescopes. The MeerKAT profile was made from a total of $\sim32.6$ h of observations with L-band receiver centred at 1284 MHz with a bandwidth of 856 MHz. The Parkes profile was made from a total of $\sim14.9$ h of observations with the two lower sub-bands of the UWL receiver with a centre frequency of 1536 MHz and a bandwidth of 1664 MHz. The profiles have been bin-scrunched down to 512 bins across the pulse phase in order to increase the S/N ratio per phase bin. Bottom panels show the dynamic spectra, made from 8 and 16 contiguous frequency bands for MeerKAT and Parkes respectively. The number of bands was chosen to result in the same frequency width per band for both telescopes and correspond to flux measurements shown in Fig.~\ref{figure:spectral_index}. The frequency range of the dynamic spectrum from the MeerKAT has been aligned to that of Parkes for the ease of comparison. The frequency evolution of the pulse profile and the steep spectral index is clearly visible.}
 \label{figure:uwl_mkt_profiles}
\end{figure*}

\subsection{Parkes observations}

Following its discovery in late 2012 \citep{CamiloEtAl2015}, search mode observations of the pulsar were undertaken with the Parkes 20-cm multibeam receiver \citep{Staveley-SmithEtAl1996} using at first the $2 \, \times \, 512 \, \times \, 0.5$ MHz Analogue Filterbank backend (AFB). This back-end was used until June 2015. In 2013 August, we started using the Digital FilterBank (DFB) versions 3 and 4 to obtain a preliminary timing solution. These observations were described by \cite{CamiloEtAl2015}. From mid-2015 until late 2016, folded mode observations were obtained with the same receiver, but with the CASPER Parkes Swinburne Recorder (CASPSR) backend. CASPSR operates at a centre frequency of 1382 MHz with a usable bandwidth of 340 MHz and is capable of performing real-time coherent dedispersion at the dispersion measure of the pulsar before folding at its topocentric period. \cite{ManchesterEtAl2013} and \citet{VenkatramanKrishnanPhDThesis} provide more information on the DFB and CASPSR backends respectively. The same receiver-backend set up was used for observations between December 2019 and January 2020 to obtain overlap between Parkes and MeerKAT data sets for better measurement of the timing jump between the two data sets. This was performed as a part of the project P1032 (PI: Venkatraman Krishnan), a project focused on obtaining complementary data to MeerKAT's relativistic binary programme (see \citealt{ksv+21} for more details).

As part of P1032, we also performed a total of $\sim 14$ hours of observations with the new UWL receiver at Parkes \citep{HobbsEtAl2020}. The data recording was performed using the \textsc{medusa} backend that records coherently dedispersed fold-mode data centred at 2368 MHz with a bandwidth of 3328 MHz. Due to the pulsar's steep spectrum  (see Section \ref{section:profile_analysis}), we were able to get useful data only from the lower 1664 MHz of the band. For the same reason, we find that the ToAs at the bottom 512 MHz of the Parkes UWL receiver provides times of arrivals that are $\sim 8 \times$ better than the previous multibeam data. An integrated profile from the UWL receiver can be seen in Fig.~\ref{figure:uwl_mkt_profiles}.

\subsection{MeerKAT observations}

The pulsar observing set up at MeerKAT is explained in detail by \cite{BailesEtAl2020}, while the details on polarisation and flux calibration are outlined in \cite{sjk+21} and \cite{SpiewakEtAl2021} respectively. All timing observations were performed with the L-band receiver under two sub-themes of MeerTime: The aforementioned RelBin \citep{ksv+21}, and the MeerKAT census of southern millisecond pulsars \citep{SpiewakEtAl2021}. MSP census observations were short ($\sim5$ min) while the RelBin observations ranged from 2048 seconds to 4 hours depending upon the orbital phase, for the necessary orbital coverage. The data presented here are from March 2019 to September 2021, and amounts to a total of $\sim32.6$ hours.

The quality of the MeerKAT L-band profiles are remarkable. The timing precision, per unit time, is $> 12\times$ better than the earlier Parkes timing. This improvement is larger than expected given the MeerKAT's $\sim6$-fold improvement in sensitivity and $\sim2$-fold increase in bandwidth compared to the Parkes multi-beam system. The steep spectral index of the pulsar measured with the combined MeerKAT + UWL dataset (see Section \ref{subsec:integrated_profile}) is a likely explanation for this this disparity as the MeerKAT usable L-band frequency goes as low as 900 MHz; even its central frequency of 1284 MHz is 100 MHz lower than that of the Parkes multibeam data sets. The pulsar's pulse profile with the MeerKAT L-band receiver is presented in Fig.~ \ref{figure:uwl_mkt_profiles}.

\section{Profile analysis}
\label{section:profile_analysis}

In this section, we report our analysis of the pulsar's flux density, spectral index, polarisation and pulse broadening due to interstellar scattering using the Parkes UWL and MeerKAT L-band datasets. Unless otherwise specified, all the analyses were performed on the integrated profile that is obtained by summing up all the observations of the pulsar per backend. This includes a total of 32.6 hours and 14.9 hours for MeerKAT L-band and Parkes UWL data respectively.

\subsection{Flux density spectrum} 

\begin{figure}
 \centering
 \includegraphics[width=\columnwidth]{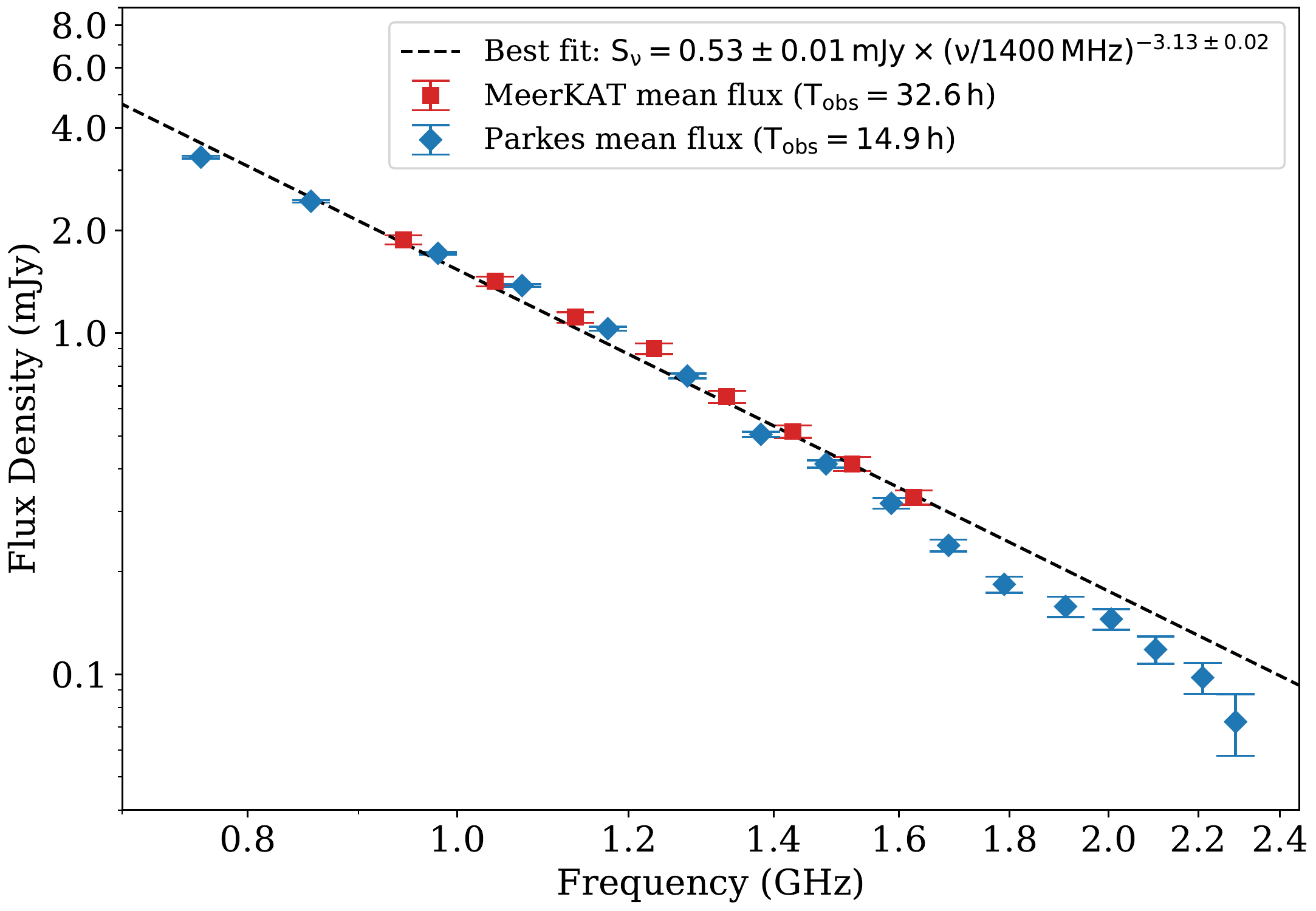}
 \caption{Mean flux density measurements for the observations performed with MeerKAT L-band (red squares) and Parkes UWL receivers (blue diamonds) with error bars indicating nominal $1 \sigma$ uncertainties, fit with power law model (black dashed line) to determine $\rm S_{\rm 1400}$ and the spectral index, $\rm \alpha$. The number of measurement points for each telescope has been chosen to result in equal bandwidth.}
 \label{figure:spectral_index}
\end{figure}

The flux density calibration of the Parkes UWL receiver was performed using observations of the Hydra A radio galaxy as well as standard pulsar reference pointings utilising pulsed source of noise (i.e. noise diode) in the ULW receiver. In the next step, a standard flux calibration technique utilising a combination of \textsc{psrchive} programs, i.e. \textsc{fluxcal} and \textsc{pac} was performed as is described in detail in \cite{2012ART....9..237V}. We have decided to divide observing bands of both telescopes such that fractional per-band frequency would be the same, thus simplifying the fitting procedure. In order to estimate the flux densities for each frequency band we have created an analytical pulse profile using the \textsc{paas} program. Subsequently \textsc{psrflux} program from the same package was used to cross-correlate it with each band's profile. The uncertainties of flux densities in each of the frequency bands were estimated by an algorithm that robustly estimates off-pulse baseline and is part of the \textsc{psrchive} package. The flux density measurements of MeerKAT data was obtained using a scaling relation from the radiometer equation as explained in \citep{SpiewakEtAl2021}.

Fig.~\ref{figure:spectral_index} presents flux density measurements made with both telescopes, as well as best fit of a power-law model, $S = S_{1400}\,(\nu/1400\,\textrm{MHz})^\alpha$. The flux density spectrum of the MeerKAT L-band data  (Fig.~\ref{figure:spectral_index}) is found to be well fit by a steep spectral index, $\alpha$ of $-3.13 \pm 0.02$, providing a mean flux density of $0.53 \pm 0.01$\,mJy at 1400 MHz. We note that in the region where data points from MeerKAT and Parkes UWL overlap, a slight offset between the points can be seen, with MeerKAT data points being slightly above those of Parkes UWL. We deduce this is due to MeerKAT antenna system temperature that was assumed in the flux density measurement method mentioned above and varying number of antennas used per observation that were integrated into the average profile used in this analysis. Additionally, we note a slight deviation from the best fit line seen for the Parkes UWL data points extending outside frequency overlap. We conclude that this effect could be due to the \textsc{psrflux} underestimating flux in the frequency bands where the pulsar signal-to-noise (S/N) ratio is low. This is especially seen in the frequency resolved plot of Fig.~\ref{figure:prof} for the Parkes UWL data at frequencies above 1.6 GHz.

\subsection{Polarisation properties}
\label{subsec:integrated_profile}

\begin{figure}
 \centering
 \includegraphics[width=\columnwidth, angle=0]{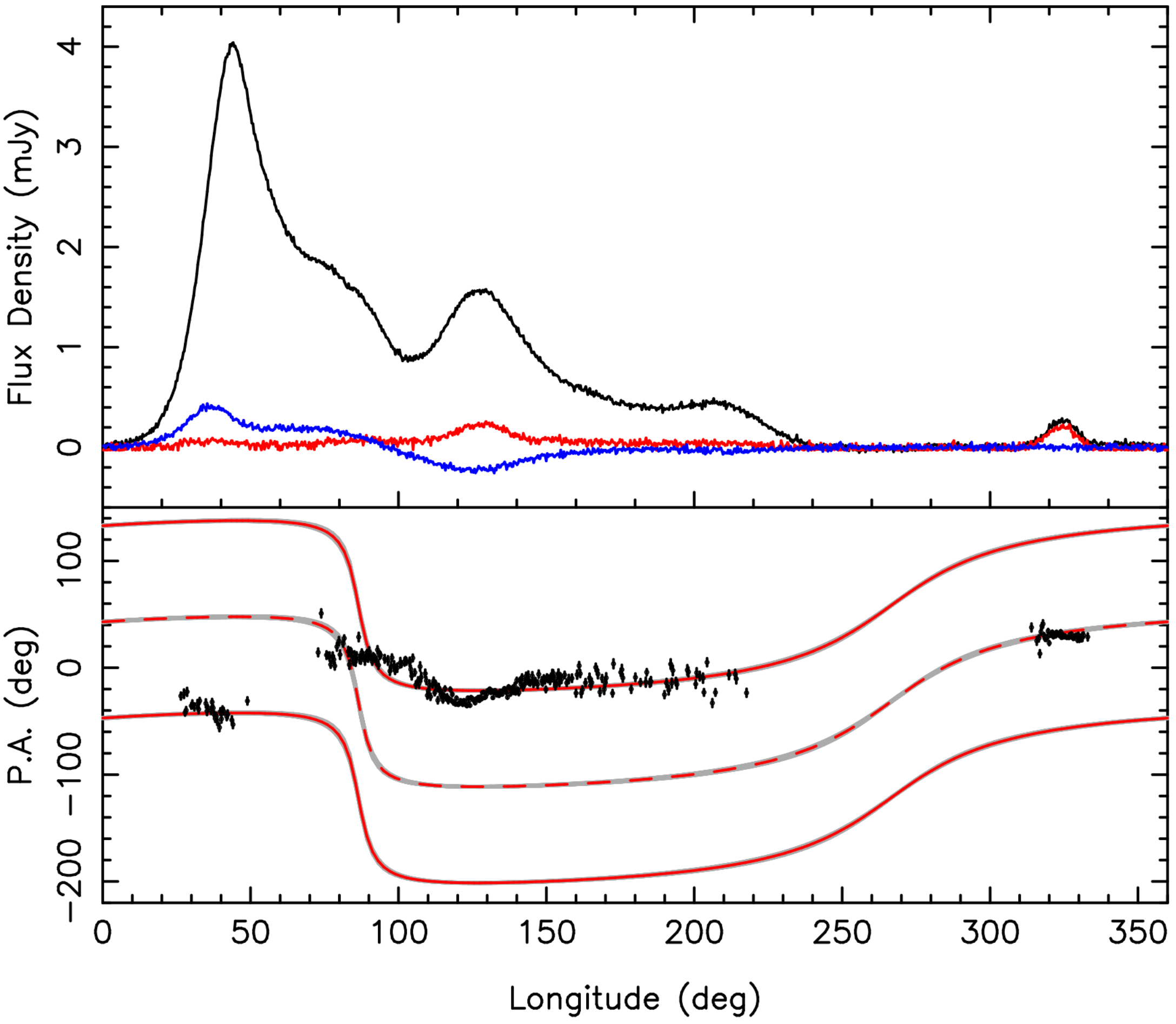}
 \caption{
 Polarisation profile as observed with MeerKAT at a central frequency of 1284 MHz, integrating a total of 32.6 hours. The top panel shows the total intensity (black), the linear (red) and circular polarisation (blue) intensity. The bottom panel shows values of the position angle (PA) swing. A Rotating Vector Model has been fitted to the PA values as shown as a red solid line and repeated offset by $180\, \deg$, while the dashed line indicates the RVM solution separated by $90\, \deg$ and intended to fit the interpulse. The grey band indicates the derived uncertainties in the determined RVM description. See Section~\ref{sec:rvm} for details.}
 \label{figure:prof}
\end{figure}

Fig.~\ref{figure:prof} presents a flux- and polarisation-calibrated average profile for \psr\, at L-band using MeerKAT. The profile is created from integrating a total of 32.6 hours and the full MeerKAT observing band (for these observations $\sim 775$\,MHz) after radio frequency interference removal. 

The pulsar shows a wide profile shape with broad multiple component features and a duty cycle in excess of 80\% at L-band. The small-amplitude component trailing the main pulse by $\sim100$ deg of longitude (called ``post-cursor''; or preceding it by about $\sim80$ deg, then called ``pre-cursor'') is not unusual for recycled pulsars but is a typical feature that distinguishes the emission from millisecond pulsars from that of ``normal'' pulsars \citep{kxl+98}. It may originate from the magnetic pole opposite to the one responsible for the main pulse (as it appears to be emitted in an polarisation mode orthogonal to the main pulse, cf.~\ref{sec:rvm}) but it may also come from a different location more generally.

The wide main pulse shows only a very small degree of polarisation, both for the linearly polarised (red) and circularly polarised (blue) component. The fact that the degree of circularly polarisation exceeds that of the linear component is uncommon for normal pulsar but is not atypical for millisecond pulsars \citep{xkj+98}. In contrast, the pre/post-cursor feature is nearly completely linearly polarised with non-detectable circular polarisation. We have calculated the phase-averaged linear polarisation fraction to be $\mathrm{L / I} = 6.2 \pm 1.6 \%$.

In the main pulse, the circular polarisation shows a sense reversal, which in normal pulsars is usually identified with the longitude of the magnetic axis \citep{Lorimer&Kramer2005}. The phase-averaged absolute circular polarisation fraction is $\mathrm{|V| / I} = 10.2 \pm 0.1 \%$. We will discuss the geometric interpretation of the polarisation properties, especially that of the position angle swing shown in the lower panel of Fig.~\ref{figure:prof} in Section~\ref{sec:rvm}.

\subsection{No evidence for scattering}

\begin{figure}
 \centering
 \includegraphics[width=\columnwidth, angle=0]{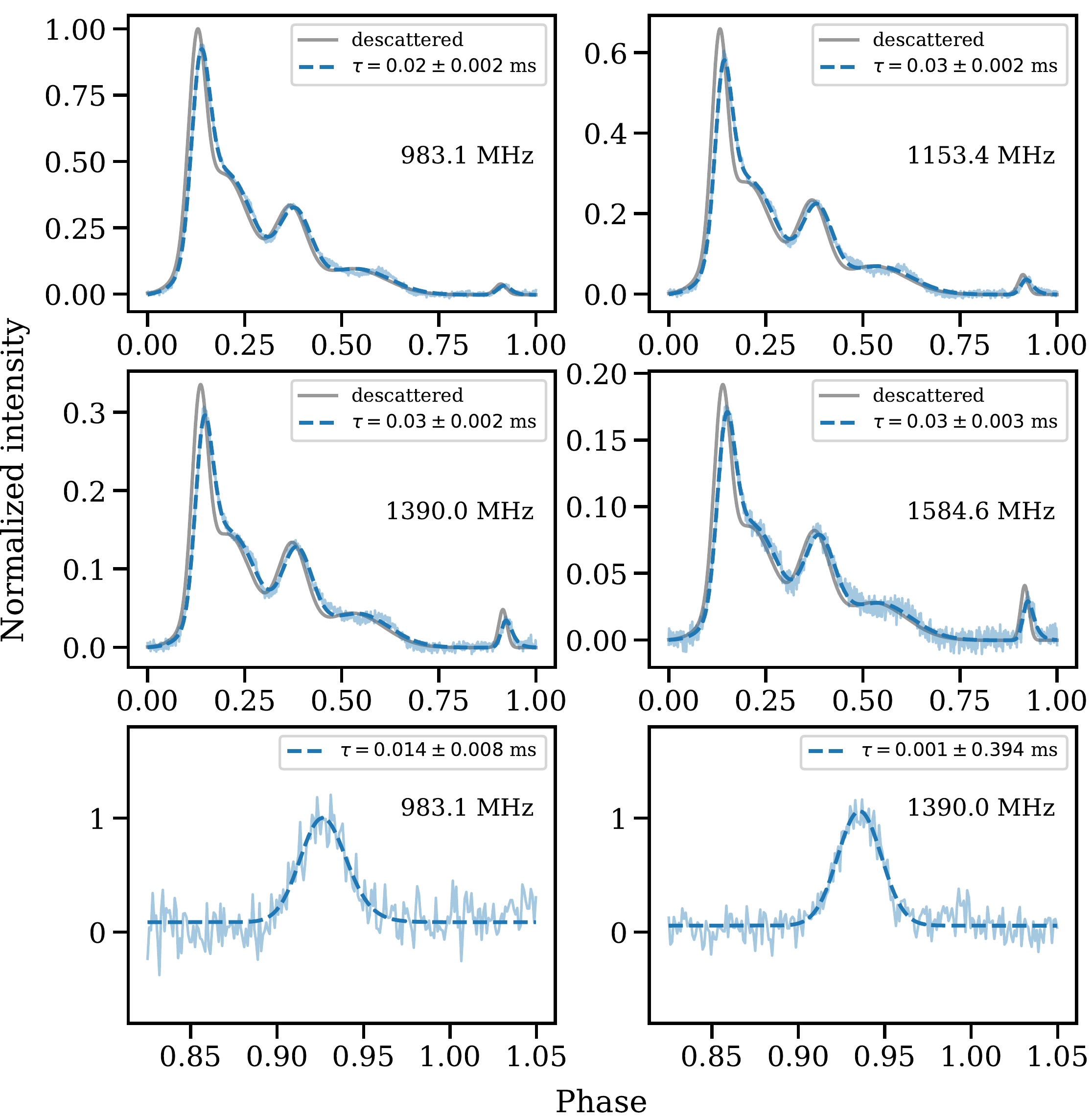}
 \caption{Top four panels: the profile shapes of PSR J0955$-$6150 across four MeerKAT L-band frequencies fitted with a scatter broadening model (blue, dashed lines). The obtained $\tau_s$ values do not show significant evolution with frequency, as is typical of ISM scattering. Bottom panels: Isolating the lone component at phase $\sim$ 0.95 we find a symmetric shape consistent with $\tau_s = 0$, further substantiating a lack of scattering broadening.}
 \label{figure:scat}
\end{figure}

We investigate evidence for scattering (i.e.~multi-path propagation) in the interstellar medium (ISM)  by fitting a pulse broadening model to the profile shapes obtained from integrating the high S/N ratio profile to four frequency channels. From these we follow two approaches. At first, the complete pulse shapes are modelled using a five-component Gaussian model (representative of the intrinsic profile), convolved with an interstellar transfer function $\propto e^{-t/\tau_s}$, where $\tau_s$ is the characteristic ISM scattering time scale \citep{Williamson1972}. The Gaussian components and $\tau_s$ values are simultaneously fit. While the channelised data are well fit by this model, as shown in Fig.~\ref{figure:scat} we observe only marginal evolution of $\tau_s$ with frequency; with best-fit $\tau_s$ values all lying between 0.02\,ms and 0.03\,ms using four channels across the band. The power law scaling, $\tau \propto \nu^{-\alpha}$, provides a flat $\alpha = -0.5 \pm 0.5$. As such $\alpha$ is poorly constrained and much less than 4 or 4.4 typically associated with simple scattering models of radio frequencies by the ionised component of the ISM (e.g. \citealt{Rickett1970} and \citealt{Rickett1977}). We note that for all our scattering fits large covariances between $\tau_s$ and many of the Gaussian component widths, used to model the underlying profile, exist. For the lowest frequency channel in Fig.~\ref{figure:scat} the anti-correlation of $\tau_s$ with the three principal Gaussian component width is > 0.8. We conclude that the obtained $\tau_s$ estimates are more likely a result of the profile's asymmetric shape and its intrinsic profile evolution, rather than due to scattering by the ISM.

As a second test, we investigated the isolated component (at phase 0.95) for evidence of scattering. The $\tau_s$ values associated with the isolated component are found to be consistent with zero as seen in the bottom panels of Fig.~\ref{figure:scat}, and are fully correlated (>0.9) with the width of this isolated component. We conclude that we do not find evidence for pulse broadening via the ISM.

We also fail to find convincing evidence for scattering in the L-band MeerKAT data of PSR~J1006$-$6311, which has similar Galactic coordinates to \psr\, (separated by 1.9$^\circ$ in Galactic longitude and 0.3$^\circ$ in latitude) and a DM of 195.99\,cm$^{-3}$\,pc \citep{DAmico98}.

The NE2001 \cite{Cordes&Lazio2002} and YMW16 \cite{YaoEtAl2017ApJ} electron density models of the Galaxy predict significantly different scattering timescales of $\tau_s^{\rm YMW16} = 0.15~\rm{ms}$ and $\tau_s^{\rm NE2001} = 0.011~\rm{ms}$\footnote{obtained using pygedm: \url{https://github.com/FRBs/pygedm}}. Correspondingly they place the pulsar at a distance of 2.17 kpc and  4.04 kpc respectively. Comparing these estimates to the results above, we note that the NE2001 model's results are more akin to our measurements.

The lack of evidence for scatter broadening of \psr's profile\, allows us to put limits on the intrinsic radio duty-cycle, which when considering the isolated component to be the inter-pulse to this pulsar, provides us with a duty-cycle > 95\%. We can also ultimately make comparisons between its intrinsic radio and the gamma ray emission, the latter of which is expected to have a wider pulse profile (more in Paper~II, in prep.). Furthermore, the apparent lack of scattering makes timing of this pulsar with the MeerKAT UHF receiver even more promising.

\section{Timing analysis}
\label{sec:timing}

\begin{table*}[h]
 \caption[]{
 \label{table:observing_details}
 Details on the observing system and the timing dataset on \psr \, used in this paper. }
 \centering
 \begin{threeparttable}
  \begin{tabular}{p{0.5in} l l l l l l l l l r}
   \hline
   \hline
   Telescope & Receiver & Backend & Centre & BW$^\star$ & nchans & CD$^*$ & Time span & \#TOAs & EFAC$^{\dagger}$ & EQUAD$^{\dagger}$  \\
       &  &  &Freq(MHz) & (MHz) &  & & (MJD) &  & &  \\
   \hline
   \multirow{5}{*}{Parkes} & & AFB & 1390 & 256 & 512 & No & 56277-57077 & 40 & 0.9& -6\\
   &20-cm&PDFB3 & 1369 & 256 & 256 & No & 56505-56943 & 23 & 0.8 & -9.0\\
   &multibeam&PDFB4 & 1369 & 256 & 512 & No & 56943-57620 & 50 & 0.75 & -5.15\\
   &&CASPSR & 1382 & 340 & 512 & Yes & 57181-58855 & 50 &0.95 & -6.59\\
   \cline{2-11}
   & Ultra-Wide-  & MEDUSA& 1536 & 1664 & 1664 & Yes & 58760-58996 & 192 & 1.5& -5.11\\
   & band Low &&&&&&&&& \\
   \hline
   MeerKAT & L-band & PTUSE & 1283.582 & 775.75 & 928 & Yes & 58568-59358 & 832 & 1.05& -6.16\\
   \hline
   \hline
  \end{tabular}
  \begin{tablenotes}
    $^\star$ Effective usable bandwidth. \\
    $^*$ Intra-channel coherent dedispersion.\\
    $^{\dagger}$ EFAC and EQUAD follows \textsc{temponest} definitions \citep{LentatiEtAl2014}.
   \end{tablenotes}
 \end{threeparttable}
\end{table*}

\subsection{Data reduction}

The data reduction for pulsar timing used standard pulsar timing analysis techniques using the \textsc{psrchive} \citep{HotanEtAl2004,2012ART....9..237V} software package and all the commands/programs specified in this section are part of this package unless explicitly mentioned/cited otherwise. We used the initial analogue filterbank (AFB) data as-is from the discovery paper of PSR J0955$-$6150 \citep{CamiloEtAl2015}. All other Parkes multibeam data were first manually mitigated of radio frequency interference (RFI) using \textsc{pazi} and \textsc{psrzap} and were scrunched to 5-minute integrations. These were polarisation calibrated using observations of a noise diode that was performed before every pulsar observation. The noise diode injects a square wave signal cycled at 11.123 Hz at $45$ degrees to each of the orthogonal signal probes. This is used to measure and compensate for the differential gain and phase that is induced across the two polarisations. The polarisation calibrated data were further scrunched such that there is one integration per observation, and two channels across the band. The data reduction for Parkes UWL was similar except that the final products had 0.7-hr integrations and 13 channels across the band.

The reduction of MeerKAT data used the \textsc{meerpipe} pipeline that performs RFI excision using a modified version of \textsc{coastguard} \citep{LazarusEtAl2016}, performs polarisation and flux calibration, and produces decimated data products that can be readily used for timing. Depending on the observing time (which in turn depended on which Meertime ``theme'' it belonged to, and what the orbital phase was), the final data product contained time integrations from 300 to 2048 seconds, and 8 channels across the observing bandwidth.

High S/N ratio observations were summed on a per backend basis to obtain a good frequency resolved pulse profile. For every backend, 2D-analytical templates were obtained by iteratively running the \textsc{paas} command for every channel from these high S/N profiles. The resultant analytical templates were then used to obtain frequency resolved time of arrivals (TOAs) using the \textsc{pat} command. More information on the observing system and the data set is given in Table \ref{table:observing_details}.

\begin{table*}
\caption{Timing parameters for \psr, obtained from the {\sc tempo2} timing package using the DDH binary model. In this and the following table, all uncertainties in the measured values are 68.3\,\% confidence limits.
}
\centering 
\begin{tabular} {l c}
\hline
\hline
\multicolumn{2}{c}{Observation and data reduction parameters}\\
\hline
Solar System ephemeris\dotfill & DE436 \\
Timescale \dotfill & TCB \\
Reference epoch for period, position and DM (MJD)\dotfill & 56983.0167959\\
First observation (MJD)\dotfill & 56277 \\
Last observation (MJD)\dotfill & 59358 \\
Solar wind electron number density, $n_{0}$ (cm$^{-3}$)\dotfill & 10.0 \\
\hline
\multicolumn{2}{c}{Spin and astrometric parameters}\\
\hline
Right ascension, $\alpha$ (J2000, h:m:s)\dotfill & 09:55:20.84737(9) \\
Declination, $\delta$ (J2000, d:m:s)\dotfill & $-$61:50:16.8945(6) \\
Proper motion in $\alpha$, $\mu_{\alpha}$ (mas\,yr$^{-1}$)\dotfill & 0.2(1)  \\
Proper motion in $\delta$, $\mu_{\delta}$ (mas\,yr$^{-1}$)\dotfill  & $-$0.1(1) \\
Spin frequency, $\nu$ (Hz)\dotfill & 500.15992019837(8) \\
Spin-down rate, $\dot{\nu}$ ($10^{-15}$\,Hz\,s$^{-1}$)\dotfill & $-$3.5663(4) \\
Dispersion measure, DM (cm$^{-3}$\,pc)\dotfill  &  160.918(8) \\
First Derivative of DM, DM1 ($10^{-3}$\,cm$^{-3}$\,pc\,yr$^{-1}$)\dotfill  & $-$6(2) \\
Second Derivative of DM, DM2 ($10^{-4}$\,cm$^{-3}$\,pc\,yr$^{-2}$)\dotfill  &11(5) \\
Rotation measure, RM (rad\,m$^{-2}$)\dotfill  & $-48(5)^{\textit{a}}$ \\
\hline
\multicolumn{2}{c}{Derived parameters}\\
\hline
Galactic longitude, $l$ ($^{\circ}$)\dotfill  & 283.684986  \\
Galactic latitude, $b$ ($^{\circ}$)\dotfill  & $-$5.737093 \\
Total proper motion, $\mu_{\text{T}}$ (mas\,yr$^{-1}$)\dotfill  & 0.2(1) \\
Position angle of proper motion, J2000, $\Theta_{\mu}$ ($^{\circ}$) \dotfill &  171(5) \\
Position angle of proper motion, Galactic, $\Theta_{\mu}$ ($^{\circ}$) \dotfill & 133(5) \\
DM-derived distance (NE2001), $d$ (kpc)\dotfill & 4.04$^{\textit{b}}$  \\
DM-derived distance (YMW16), $d$ (kpc)\dotfill & 2.17$^{\textit{b}}$  \\
Parallax, $\bar{\omega}$ (mas)\dotfill  & 0.24$^{\textit{c}}$ \\
Galactic height, $z$ (kpc)\dotfill & $-0.40(6)^{\textit{c}}$  \\
Heliocentric transverse velocity, $v_{\text{T}}$ (km\,s$^{-1}$)\dotfill & $13(2)^{\textit{c}}$  \\
Spin period, $P_{0}$ (ms)\dotfill & 1.9993605237367(2)  \\
Spin period derivative, $\dot{P}$ ($10^{-20}$\,s\,s$^{-1}$)\dotfill & 1.42601(17) \\
Total kinematic contribution to period derivative, $\dot{P}_{\text{k}}$ ($10^{-20}$\,s\,s$^{-1}$)\dotfill & $-0.075$  \\
Intrinsic spin period derivative, $\dot{P}$ ($10^{-19}$\,s\,s$^{-1}$)\dotfill & 1.501(9) \\
Surface magnetic field strength, $B_{\text{surf}}$ ($10^{9}$\,G)\dotfill  & 0.17 \\
Characteristic age, $\tau_{\text{c}}$ (Gyr)\dotfill & 2.1  \\
Spin-down power, $\dot{E}$ ($10^{34}$\,erg\,s$^{-1}$)\dotfill & 7.4  \\
\hline
\hline
\multicolumn{2}{l}{$^{\textit{a}}$ Obtained using \textsc{rmfit} program in the \textsc{psrchive} software package.}\\
\multicolumn{2}{l}{$^{\textit{b}}$ Assuming a $20\%$ uncertainty in the distance}\\
\multicolumn{2}{l}{$^{\textit{c}}$ Assuming DM-derived distance (NE2001)}\\
\label{tab:timing}
\end{tabular}
\end{table*}

\begin{table*}
\caption{Binary parameters measured for PSR\,J0955$-$6160 obtained using \textsc{tempo2}. Square brackets indicate derived quantities. For the DDGR solution, the $\dot{P}_{\rm b}$ is fitted as a term in addition to the (very small) orbital decay caused by the emission of gravitational waves.
For the grid solution, the values for the $\chi^2$ correspond to those of the best point in the grid, they are slightly lower than the DDGR solution because for each point in the grid there are two parameters ($M$ and $M_{\text{c}}$) that are assumed, not fitted. }
\centering 
\begin{tabular} {l c c c }
\hline
\hline
Binary model\dotfill & DDGR & DDH  & DDGR $\chi^{2}$ grid \\[0.5ex]
Number of ToAs \dotfill & 1186 & 1186 & 1186 \\
weighted rms of ToA residuals ($\mu s$) \dotfill & 2.02 & 2.02 & 2.02 \\
$\chi^2$ of fit \dotfill & 1197.82 & 1199.91 & 1196.11 \\
$\chi^2$ / number of degrees of freedom\dotfill & 1.009 & 1.011 & 1.009 \\
\hline
\multicolumn{4}{c}{Keplerian orbital parameters}\\[0.5ex]
\hline
Orbital period, $P_{\text{b}}$ (days)\dotfill & 24.57839502(6) & 24.57839502(6) & - \\[0.5ex]
Projected semi-major axis of the pulsar orbit, $x$ (lt-s)\dotfill & 13.282477(2) & 13.2824767(6) & - \\[0.5ex]
Epoch of periastron, $T_{0}$ (MJD)\dotfill & 56287.604348(6) & 56287.604349(6) & - \\[0.5ex]
Orbital eccentricity, $e$\dotfill & 0.11750575(1) & 0.11750575(1) & - \\[0.5ex]
Longitude of periastron at $T_{0}$, $\omega$ ($^{\circ}$)\dotfill &  202.92940(9) & 202.92941(9) & - \\[0.5ex]
\hline
\multicolumn{4}{c}{Post-Keplerian orbital parameters}\\[0.5ex]
\hline
Rate of advance of periastron, $\dot{\omega}$ ($^{\circ}$\,yr$^{-1}$)\dotfill & [0.0014809] & 0.00152(1) & - \\[0.5ex]
Einstein delay, $\gamma$ (ms) \dotfill & [0.5362] & [0.5417] & - \\
(Excess) Orbital period derivative, $\dot{P_{\text{b}}}$ ($10^{-12}$\,s\,s$^{-1}$)\dotfill & 11(7) & 11(7) & - \\[0.5ex]
Orthometric amplitude of Shapiro delay, $h_{3}$ ($\mu$s)\dotfill & - & 0.89(7) & - \\[0.5ex]
Orthometric ratio of Shapiro delay, $\varsigma$\dotfill & - & 0.88(2) & - \\[0.5ex]
\hline
\multicolumn{4}{c}{Mass and inclination measurements}\\
\hline
Mass function, $f$ (M$_{\odot}$)\dotfill & 0.004164980(1) & 0.0041649796(5) & - \\[0.5ex]
Total mass, $M$ (M$_{\odot}$)\dotfill & 1.96(2) & [1.9602] & 1.96(3) \\[0.5ex]
Pulsar mass, $M_{\text{p}}$ (M$_{\odot}$)\dotfill & 1.71(3) & - & 1.71(2) \\[0.5ex]
Companion mass, $M_{\text{c}}$ (M$_{\odot}$)\dotfill & 0.254(2) & - & 0.254(2) \\[0.5ex]
Orbital inclination, $i$ ($\deg$)\dotfill & - & - & $83.2(4)$ \\[0.5ex]
\hline
\hline
\label{tab:binary}
\end{tabular}
\end{table*}

\begin{figure*}
 \centering
 \includegraphics[width=0.8 \textwidth]{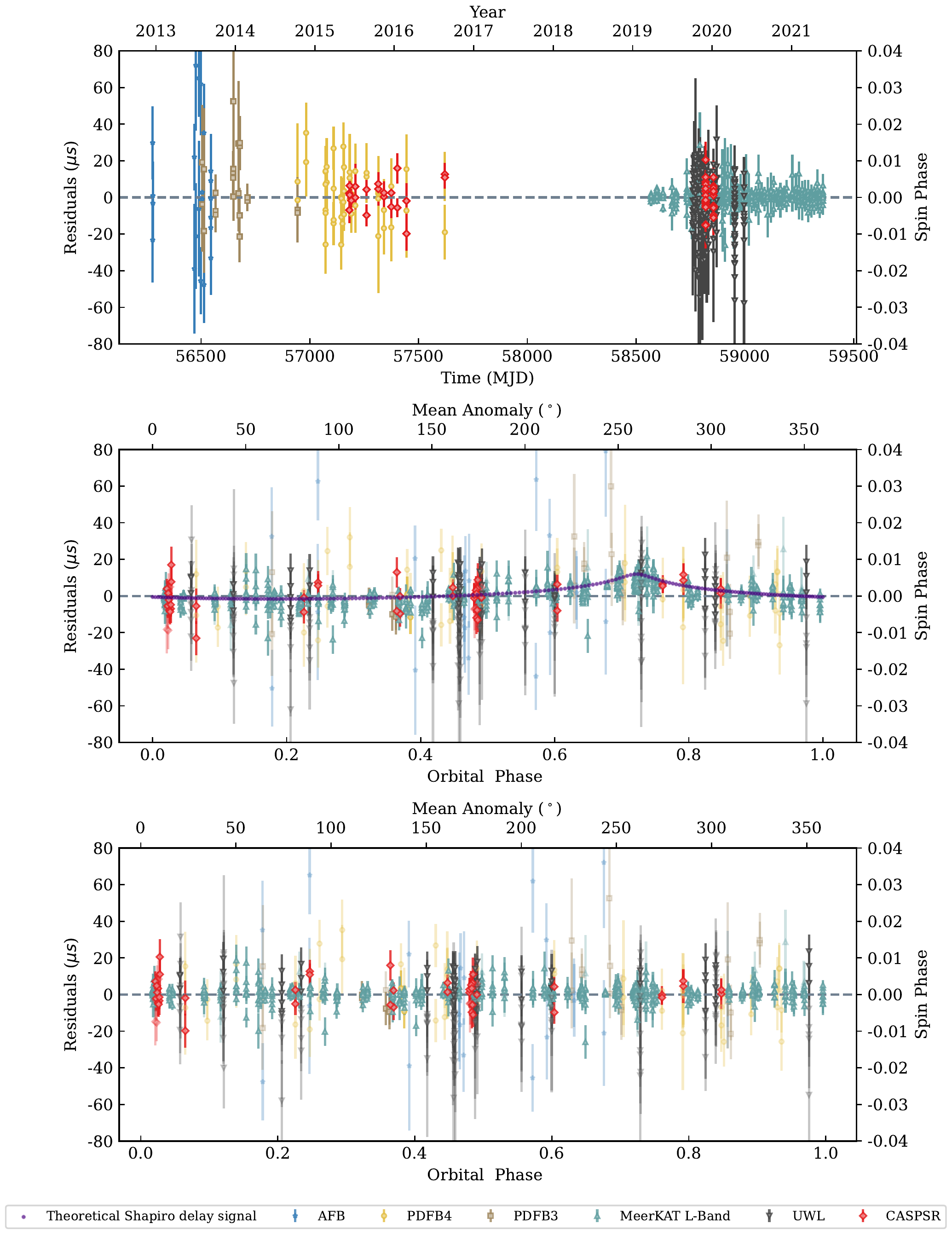}
 \caption{Top to bottom: Post-fit residuals for the timing of \psr, obtained with the {\sc tempo2} DDH solution as a function of (1) time, (2) orbital phase without subtracting the full Shapiro delay signal and (3) orbital phase after subtracting the Shapiro delay signal. The orbital phase is measured from periastron; the superior conjunction happens at a mean anomaly of 259.9 degrees. The colours denote the different telescopes and back-ends used in the timing. {\em Blue}: Parkes Analogue Filterbank (AFB) data, {\em Brown and Yellow}: Parkes Digital Filterbank (DFB) data, {\em Red}: CASPER Parkes Swinburne Recorder (CASPSR) data, which was taken both before 2017 and after 2019, thus establishing timing continuity for the whole data set, {\em Gray}: Parkes UWL data and {\em Teal}: MeerKAT L-band data. The middle plot also shows the theoretical Shapiro delay signal for the best value of the orbital inclination. In the bottom two plots, TOAs with precision worse than 10$\mu$s are made semi-transparent for clarity.}
 \label{figure:residuals}
\end{figure*}

\subsection{Timing}

For the TOA analysis we used the {\sc tempo2} pulsar timing package \citep{HobbsEtAl2006} and {\sc temponest}, a bayesian parameter estimation plugin to {\sc tempo2} that also facilitates fits for, among other things,  power law models for red and DM noise in the data \citep{LentatiEtAl2014}.

To describe the telescope's motion relative to the Solar System Barycentre, we used JPL's DE436 Solar System ephemeris. All ToAs were transferred to Universal Coordinated Time (UTC) and then to the terrestrial time standard, ``TT(TAI)" that is derived from the ``International Atomic Time" timescale. To describe the pulsar's orbital motion, we use two models related to the theory-independent model of \cite{DD2} (henceforth designated as "DD"). The DD model describes the orbital motion as being essential a Keplerian orbit with small relativistic perturbations. With pulsar timing, we can only measure five of its elements: the orbital period ($P_{\rm b}$), orbital eccentricity ($e$), the semi-major axis of the pulsar's orbit projected along the ling of sight ($x$), the longitude of periastron ($\omega$) and the time of passage of periastron ($T_0$). The relativistic perturbations are quantified, in a general and theory independent way \citep{DamourTaylor1992}, by the so-called ``Post-Keplerian'' (PK) parameters, which are: rate of advance of periastron ($\dot{\omega}$), the variation of the orbital period ($\dot{P}_{\rm b}$, which includes the orbital decay caused by the emission of gravitational waves) and the Einstein delay ($\gamma$, which is caused by the orbital variation of the special relativistic time dilation and general relativistic gravitational redshift). In addition, the model includes the aforementioned Shapiro delay.

The first model we used is the DDGR orbital model, which assumes the validity of GR to describe all relativistic effects and fits directly for the masses of the two objects in the system. In parallel, we used the theory-independent orbital model (DDH) in order to understand which relativistic effects are effectively being measured; this is important for verifying whether they are all consistent with each other within the framework of GR. This model is nearly identical to the DD model; the only difference is the PK parameters used to describe the Shapiro delay: in the DD model these are the ``range'' ($r$) and ``shape'' ($s$) parameters, in the DDH model these are the orthometric amplitude ($h_3$) and the orthometric ratio ($\varsigma$,  \citealt{2010MNRAS.409..199F}). The advantage of using $h_3$ and $\varsigma$ is that they have, particularly for lower inclinations, a much lower correlation between themselves than $r$ and $s$; hence, they provide a better description of mass and inclination constraints introduced by the Shapiro delay. For orbital inclinations close to $90^\circ$, the two models are equivalent.

\begin{figure*}
 \centering
 \includegraphics[width=\textwidth]{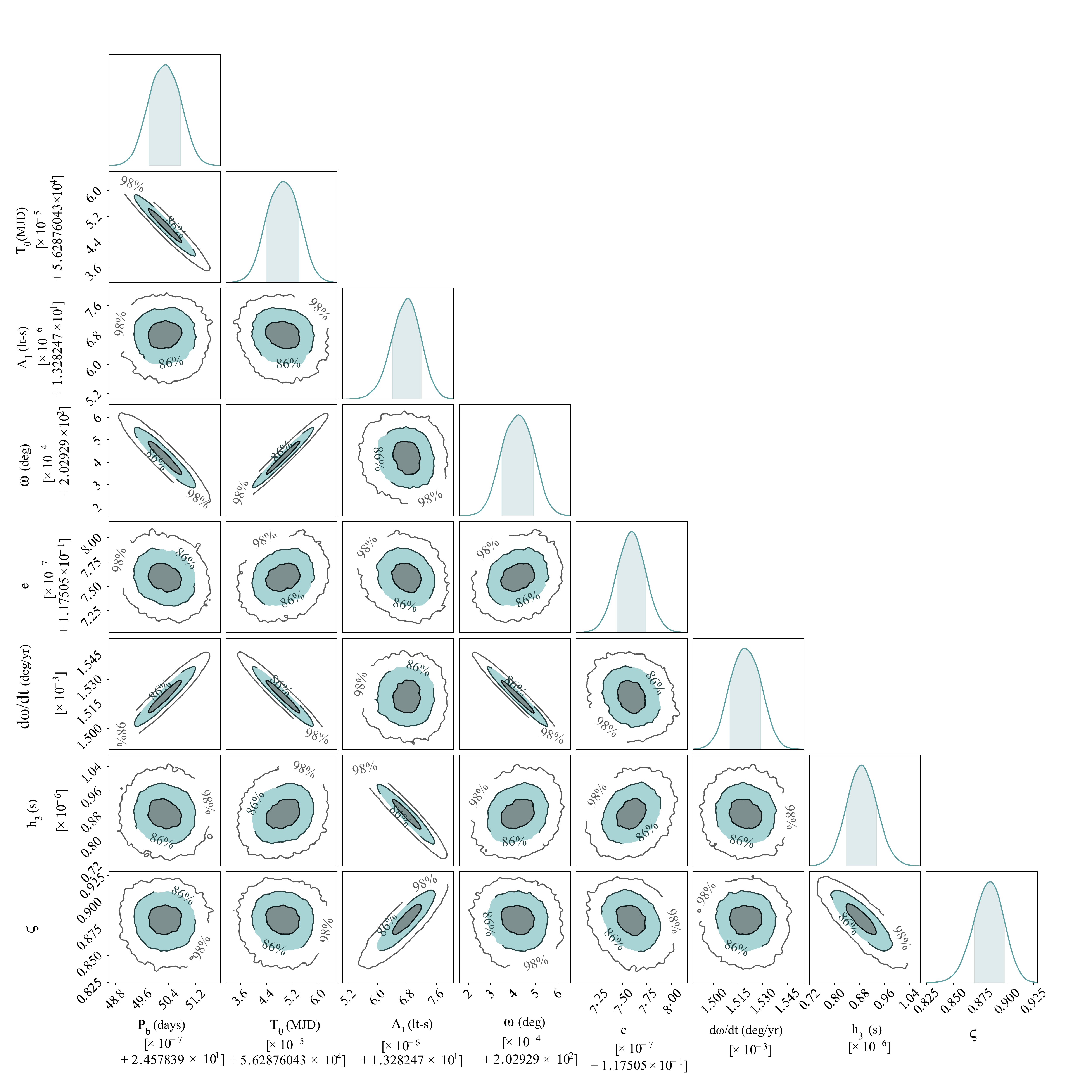}
 \caption{ A corner plot showing the posterior distributions of the orbital and Post-Keplerian parameters and the correlations between them for PSR J0955-6150 obtained from the non-linear timing of the pulsar  with the DDH binary model using \textsc{temponest}. The off-diagonal elements show the correlation between the parameters and are marked contours that define 39\%, 86\% and 98\% C. L. while the diagonal elements show the marginalised 1D posterior distributions of the parameters with the shaded region marking nominal 1$\sigma$ or 68.27\% of the probability.\label{fig:corner}}
\end{figure*}

For both DDH and DDGR timing models, we performed Bayesian non-linear fits of the timing model to our data using \textsc{temponest}. Apart from the timing parameters, we fit for white noise parameters; EFAC and EQUAD, per backend that modify the formal TOA uncertainties, and a power law DM noise model as described in \cite{LentatiEtAl2014}. We also performed fits for a red timing noise model, but the posteriors indicated that the red noise in the data set is negligible. Hence we ignored red timing noise for the rest of our analysis. The estimates of the pulsar parameters can be found in Table \ref{tab:timing} and \ref{tab:binary}, the uncertainties on the parameters represent 68.3\% confidence levels that are scaled to a reduced $\chi^2$ of 1. The first table has the spin and astrometric parameters for the pulsar; the second has the binary parameters derived according to the DDGR, DDH models, and the results of our Bayesian analysis of the masses of the components using a $\chi^2$-grid, which is described in section~\ref{section:chi2map}. The TOA residuals (i.e., the TOA minus the prediction of the ephemeris for that rotation) are depicted in Figure~\ref{figure:residuals}. Figure \ref{fig:corner} shows the marginalised 1D-posterior distributions and the 2D-correlation contours for the orbital and post-Keplerian parameters that are relevant for this paper.

In the remainder of this section we call the reader's attention to some of the timing parameters we have measured, and discuss their significance, with a special focus on the post-Keplerian parameters and the masses of the pulsar and its companion.

\subsection{Position and Proper motion}

The timing yields a very precise position of the pulsar in the sky. This allows a search for counterparts at optical
wavelengths. Inspecting the GAIA data release 3 \citep{2021A&A...649A...1G}, we find no counterparts within 3\arcsec of the position of the pulsar. However, this goes only to a magnitude of about 20. A deeper optical map of the Southern Galactic plane has been obtained by Cerro Tololo's DECam Plane Survey \citep{2018ApJS..234...39S}, where the faintest objects have magnitudes of 23.7, 22.8, 22.3, 21.9, and 21.0 mag (AB) in the grizY bands, respectively, and average seeing of about 1\arcsec. Again, no clear counterparts can be detected within 3\arcsec of the position of the pulsar. In the direction of this pulsar, the extinction is 1.15, 0.773, 0.567, 0.432 and 0.376 magnitudes for the g, r, i, z and Y bands, respectively \citep{2011ApJ...737..103S}. This implies that the companion WD must be fainter (in the g band) than magnitude $\sim$21.1. This is not surprising: the WD companion of the eccentric MSP PSR~J2234+0611 has a g magnitude of 22.17 \citep{2016ApJ...830...36A}. Furthermore, PSR~J2234+0611 is at a distance $0.95 \pm 0.04$ kpc \citep{2019ApJ...870...74S}; the estimated distance to \psr\, is at least twice as large, which would increase its magnitude by 1.5. This means that even without extinction, the companion of PSR~J2234+0611 would not be detectable at the distance of \psr.

We have also looked for counterparts in the near-infrared VISTA Hemisphere Survey, which has a target depth is 20.6,19.8 and 18.5 magnitudes for the J, H and K bands respectively \citep{2019A&A...630A.146S}. Again, no clear counterparts are seen at the position of the pulsar. Thus, we cannot confirm that the companion is a He WD - if so, it is too faint to be detectable in current surveys.

Our measurement of the proper motion of \psr\, shows it is unusually small, and consistent with no detectable motion both in Right Ascension ($\alpha$) and Declination ($\delta$); the total proper motion $\mu$ is only 0.2(1) mas/yr. This yields a very small Heliocentric velocity: if we use the NE2001 model of the electron distribution of the Galaxy \citep{Cordes&Lazio2002}, then the distance is around 4.0(8) kpc (after assuming a distance uncertainty of about 20 \%) and the resulting heliocentric velocity is $3.8\, \pm \, 2.5 \, \rm km \, s^{-1}$; using the YMW16 model \citep{YaoEtAl2017ApJ} we obtain a distance of 2.2(4) kpc and a heliocentric velocity of $2.0\, \pm \, 1.3 \, \rm km \, s^{-1}$ (these estimates assume only the uncertainty in the proper motion, which is, in relative terms, much larger than the uncertainty in the distance). This is very small compared to the average velocities of MSPs (e.g., \citealt{2011ApJ...743..102G,2016MNRAS.458.3341D,2018ApJS..235...37A}), and even smaller compared to the velocities of normal pulsars \citep{2005MNRAS.360..974H}.

However, the velocity of the pulsar relative to its local standard of rest is much larger. Following the simple method described by \cite{2019ApJ...881..165Z}, we obtain peculiar velocities of $\sim 133 \, \rm km \, s^{-1}$ and $\sim 78 \, \rm km \, s^{-1}$ for the two distances listed above. These are nearly parallel to the Galactic plane: for instance, for the NE2001 distance, the perpendicular velocity is only $\sim 8 \, \rm km \, s^{-1}$. These peculiar velocity estimates are much more typical of what one finds among the general MSP population.

\subsection{Spin period derivative}
\label{sec:pdot}

The proper motion measured is important for estimating the intrinsic spin-down of the pulsar, $\dot{P}_{\rm int}$, from the observed spin-down, $\dot{P}_{\rm obs}$:
\begin{equation}
\frac{\dot{P}_{\rm int}}{P}  = \frac{\dot{P}_{\rm obs}}{P} - \frac{\mu^2 d}{c} - \frac{a}{c},
\end{equation}
where $d$ is the distance from the Earth to the system, $a$ is the difference of the accelerations of the Solar system and of the pulsar's system in the gravitational field of the Galaxy, projected along the direction from Earth to pulsar, and $c$ is the speed of light. The contribution to $\dot{P}$ that depends on $\mu$ is the Shklovskii effect \citep{1970SvA....13..562S}, for the NE2001 distance estimate this is very small, only $P \mu^2 d/c = 7.7 \, \times \, 10^{-25}$, a consequence of the small Heliocentric proper motion of the system.

The Galactic acceleration can be calculated using the analytical expressions in, e.g., \cite{2009MNRAS.400..805L}; these are sufficiently accurate given the small Galactic latitude of the pulsar. In these expressions, we used an estimate of the distance of the Solar System to the Galactic centre and rotational velocity of the Galaxy ($D = 8.275(34)\,$kpc, $v_{\rm Gal} \, = \, 240.5(41)\, \rm km \, s^{-1}$) from \cite{2021arXiv210701096A}. For the pulsar distance, we used the NE2001 estimate. From this, we get $P a/c \, = \, -0.75 \, \times \, 10^{-21}$.

Adding these two contributions, we obtain a total $\dot{P}$ correction $\dot{P}_{k} = -0.75 \, \times \, 10^{-21}$, which is completely dominated by the Galactic acceleration term. From this, we obtain an intrinsic spin-down of $1.501(9) \, \times \, 10^{-20}$, which is similar, but slightly larger, than $\dot{P}_{\rm obs}$. From this, we estimate the characteristic age, magnetic field and spin-down power values presented in Table~\ref{tab:timing}.

\subsection{Post-Keplerian parameters}

\begin{figure*}
 \centering
 \includegraphics[width=0.8\textwidth]{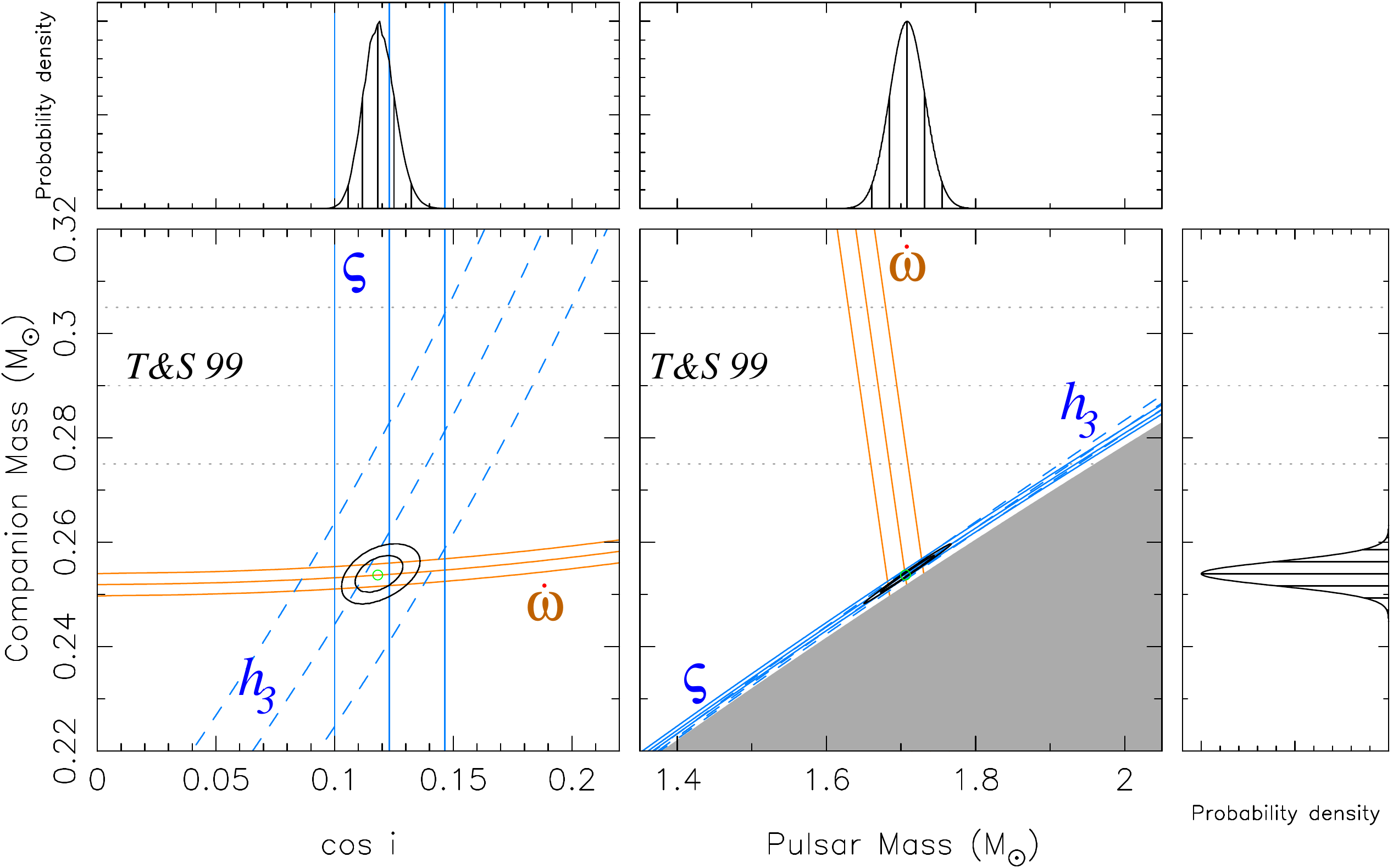}
 \caption{Mass and orbital inclination constraints for the \psr\, binary system. In both panels, the lines represent the nominal and $\pm \, 1 \, \sigma$ mass and inclination constraints derived from the PK parameters of the {\sc tempo2} DDH solution in Table~\ref{tab:binary} (orthometric amplitude of the Shapiro delay, $h_3$, in dashed blue, the orthometric ratio, $\varsigma$, in solid blue and the rate of advance of periastron, $\dot{\omega}$, in solid orange), these constraints are calculated assuming that they are purely relativistic and that GR is the correct theory of gravity. The grey dotted lines show the \cite{1999A&A...350..928T} for the $M_{\rm c}$ of this system given its orbital period. The green circles indicate the best-fit masses and orbital inclination determined with the DDGR model. {\em Left}: $\cos i$ - $M_{\rm c}$ plane.  {\em Right}: $M_{\rm p}$ - $M_{\rm c}$ plane. The grey area is excluded because $\sin i \, \leq \, 1$.}
 \label{figure:mass_mass}
\end{figure*}

As shown in Table~\ref{tab:binary}, using the DDH solution, we can measure three PK parameters, $\dot{\omega}$, $h_3$ and $\varsigma$, with high significance. The mass and inclination constraints that result from these parameters according to GR are depicted graphically in Figure~\ref{figure:mass_mass}, where each triplet of lines depicts the mass and inclination constraints derived from their nominal values and $68.3\%$ confidence level uncertainties.

\subsubsection{The rate of advance of periastron}

\label{sec:omdot}
If the rate of advance of periastron ($\dot{\omega}$) is purely relativistic, then in GR this effect yields the total mass of the system, in solar masses:
\begin{equation}
M \, = \, \frac{1}{T_\odot} \left[ \frac{\dot{\omega}}{3} (1 - e^2) \right]^{\frac{3}{2}} \left( \frac{P_{\rm b}}{2 \pi} \right)^{5/2},
\end{equation}
where $T_\odot \equiv {\cal G M_{\odot}^{\rm N}} / c^3$, with ${\cal G M_{\odot}^{\rm N}}$ being the solar mass parameter \citep{2016AJ....152...41P}. Since this and the speed of light $c$ are defined exactly, the same applies to $T_{\odot}$, which has a numerical value of $4.9254909476412675...\, \rm  \mu s$.

The constraint on the total mass of the binary that results from $\dot{\omega}$ ($M = 1.96(2) M_{\odot}$) is depicted graphically by the orange lines in Fig.~\ref{figure:mass_mass}. From this, we obtain from Kepler's third law an inclination-independent estimate of the orbital separation, 
\begin{equation}
    a = c \left[ M T_{\odot} \left( \frac{P_{\rm b}}{2 \pi}  \right)^2    \right]^{1/3} = 30.96 \, \rm Gm,
\end{equation} 
a value that we will need below. However, for binary systems with wide orbits, the observed $\dot{\omega}$ might not be purely relativistic. Generally, the second most important contribution is a geometric contribution from the proper motion, $\dot{\omega_{\mu}}$. Re-arranging the expressions in \cite{1996ApJ...467L..93K}, we obtain:
\begin{equation}
\dot{\omega}_{\mu} \, = \, \frac{\mu}{\sin i} \cos \left( \Theta_{\mu} - \Omega \right),\label{eqn:omega_mu}
\end{equation}
where $\Theta_{\mu}$ is the position angle of the proper motion and $\Omega$ is the position angle of the line of nodes. This expression is valid if we use the ``observer's reference frame'', where the position angles start from the North and increase anti-clockwise through the East. In this system, an orbital inclination smaller than $90\, \deg$ implies that the line of sight component of the orbital angular momentum points towards the Earth.

In the DDGR solution we get a nominal estimate of $i = 83.2 \deg $ or its equally likely counterpart,  $180-i = 96.8 \deg$. Luckily, the sign of $\sin i$ for both the cases (which is what is needed in \ref{eqn:omega_mu}) is positive. Based on this value and our current estimate of $\mu$, the maximum value of $\dot{\omega}_{\mu}$ is $\sim \, 5 \, \times \, 10^{-8}\, \deg\, \rm yr^{-1}$; which is uniquely small among eMSPs. This is $\sim 250$ times smaller than the current measurement uncertainty for $\dot{\omega}$.

Generally, other contributions to $\dot{\omega}$ are very small compared to $\dot{\omega}_{\mu}$, however, this is not the case in this pulsar. Using eq. (5.17) of \cite{Damour&Schafer1988}, we find that the contribution due to the Lense-Thirring effect caused by the spin of the pulsar has a maximum value of the order of:
\begin{equation}
    | \dot{\omega}_{\rm LT} | = 5.8 \times 10^{-8} \, \deg\, \rm yr^{-1},
\end{equation}
assuming the moment of inertia of the pulsar is $10^{38} \, \rm kg \, m^2$. This is very similar to our current estimate of $\dot{\omega}_{\mu}$.

All of this means that the observed $\dot{\omega}$ is purely relativistic. It also implies that a two order of magnitude improvement of the precision of $\dot{\omega}$ will result in a similar improvement in the precision of
$M$, i.e., an eventually achievable uncertainty of about $10^{-4} \, M_{\odot}$, an extraordinarily precise measurement of the mass of a MSP binary.

\subsubsection{The Shapiro delay}
\label{sec:shapiro}

One of the main results in this paper is the precise measurement of the Shapiro delay. This is only possible given the high timing precision achievable with MeerKAT and, of course, the high orbital inclination of about $83^\circ$ (or $180- 83^\circ$). In Figure~\ref{figure:mass_mass}, we can see the mass constraints introduced, according to GR, by the two Shapiro delay parameters represented by blue lines, solid for $\varsigma$ and dashed for $h_3$ (see eqs. 22 and 23 in \citealt{2010MNRAS.409..199F}).

By itself, the  Shapiro delay already yields useful mass estimates: they already characterise the WD companion as a likely He~WD and the pulsar as likely massive, however, these constraints are not particularly precise. It is when this effect is combined with the the measurement of $\dot{\omega}$, we obtain an order of magnitude improvement on the precision of the mass measurements. This and other results will be discussed more quantitatively in section~\ref{section:chi2map}.

The fact that all PK parameters cross at the same locations in the two panels of Fig.~\ref{figure:mass_mass} constitutes a successful test of GR. However, this is not of great interest given the low precision of the masses obtained via the Shapiro delay. Instead, we can think of this multiple coincidence of PK parameters as a confirmation of the basic validity of the mass-measuring method being used, and in particular as a verification of our assertion that $\dot{\omega}$ is a purely relativistic effect, i.e., it has no quantifiable classical contributions caused by e.g., the rotation of the companion.

\subsubsection{Variation of the orbital period}

For the masses determined by the DDGR model, the orbital decay caused by the emission of gravitational waves is negligible, $-3.8 \, \times \, 10^{-17}$. Much larger is the kinematic contribution to $\dot{P}_b$ ($\dot{P}_{b, \rm K}$) caused by the acceleration of the system in the gravitational field of the Galaxy (this also includes an almost negligible contribution from the Shklovskii effect). Using the same methods as those used in section~\ref{sec:pdot}, we estimate $\dot{P}_{b, \rm K}\, = \, -0.80\, \times\, 10^{-12}$ for the NE2001 distance ($d \, \sim \, 4.0\,$kpc) and $-0.44\, \times\, 10^{-12}$ for the YMW16 distance $d \, \sim \, 2.2\,$kpc.

Fitting for $\dot{P}_{b}$ in the DDH model (something not done in the solution presented in Table~\ref{tab:binary}), we obtain $\dot{P}_{b} \, = \,  9\, \pm 6 \, \, \times \, 10^{-12}$ (68.3 \% confidence limit). This is 2-$\sigma$ consistent with the much smaller expectation for the kinematic contribution to $\dot{P}_b$. This means that we will have to improve the precision of $\dot{P}_{b}$ by more than one order of magnitude in order to start detecting the kinematic contribution. This is desirable since a precise measurement of the kinematic $\dot{P}_b$ can yield a precise distance to the system \citep{1996ApJ...456L..33B,StairsEtAl1998}; it is also achievable because the precision of the measurement of $\dot{P}_b$ will improve dramatically over the next few years with continued MeerKAT timing of the pulsar, particularly with the UHF receiver. However, without the TOAs from that receiver, we cannot yet simulate how long it will take until a the measurement of $\dot{P}_{b, \rm K}$ can yield a precise distance to the pulsar. Detailed simulations including the UHF data will be published elsewhere.

\subsubsection{Variation of the projected semi-major axis and the Einstein delay}

The proper motion also produces a secular variation of the semi-major axis. Again, re-arranging the expressions in \cite{1996ApJ...467L..93K} and using the convention described in section~\ref{sec:omdot}, one obtains:
\begin{equation}
\dot{x}_{\mu}\, = \, x\, \mu \cot i \sin(\Theta_{\mu} - \Omega),
\end{equation}
which for \psr\, yields an estimate\footnote{This is assuming the nominal value of the proper motion, which is only 1.5-$\sigma$ significant} of at most $\pm \, 5 \, \times \, 10^{-17} \, \rm lt-s\, s^{-1}$. Fitting for $\dot{x}$ in the DDH model (something that was also not done in the solution presented in Table~\ref{tab:binary}), we obtain $\dot{x} \, = \, 4 \, \pm \, 6\, \times\, 10^{-15} \rm lt-s\, s^{-1}$. The uncertainty in this measurement is still $\sim 130$ times larger than the maximum value of $\dot{x}_{\mu}$.

The $\dot{x}_{\mu}$ is unusually small among eMSPs, partly because of the small proper motion, but also because of the high inclination. For this reason, we will now estimate the future ability to measure the Einstein delay, $\gamma$.

For timing baselines that are much shorter than the precession timescale of binary (which is certainly the case for \psr, where the precession timescale is $360\, \deg / \dot{\omega} \sim 0.24\,$ Myr), both $\dot{x}_{\mu}$ and $\gamma$ are hopelessly correlated. The reason for this is that, under this condition, the effect of $\gamma$ on the timing is merely to produce an additional secular, linear contribution to the observed variation of the projected semi-major axis $\dot{x}_{\rm obs}$, this is given by eq. 25 of \cite{2019MNRAS.490.3860R}:
\begin{equation}
\dot{x}_{\rm obs} \, = \, \dot{x}_{\mu} - \frac{\gamma \dot{\omega}}{\sqrt{1 - e^2}} \sin \omega.
\end{equation}
For the masses in the DDGR solution in Table~\ref{tab:binary}, GR predicts $\gamma \, = \, 0.536 \, \rm ms$. From this, we estimate that the second term on the right is $- \, 1.7 \, \times \, 10^{-17} \, \rm lt-s\, s^{-1}$, still twice as small as the maximum value for $\dot{x}_{\mu}$. Thus, although $\dot{x}_{\mu}$ is exceptionally small in this system, the effect of $\gamma$ in the timing is still smaller than that, and therefore $\gamma$ is not independently measurable - unless the proper motion proves to be much smaller than our current estimate. This superposition with the $\dot{x}$ from proper motion prevents the measurement of $\gamma$ in most eccentric, wide binary pulsars, the exception being, to date, the system studied by \cite{2019MNRAS.490.3860R}, PSR~J0514$-$4002A.

\subsection{Bayesian mass estimates assuming GR} \label{section:chi2map}

We now proceed to estimate the component masses and their uncertainties using a self-consistent Bayesian approach commonly used for this purpose (see e.g., \citealt{2002ApJ...581..509S}); this is based on the quality of the {\sc tempo2} timing fit (measured by the resulting $\chi^2$) for the relevant physical parameters we want to measure, in this case the orbital inclination and the masses.

Because the kinematic effects on $\dot{\omega}$ and $\dot{x}$ are not measurable, we have no information whatsoever on the line of nodes, $\Omega$ (see sections~\ref{sec:omdot} and \ref{sec:shapiro}). This means that, instead of mapping the $\chi^2$ for a 3-D space, with axes given by $\Omega$, $\cos i$ and $M$, as in \cite{2011MNRAS.412.2763F} and \cite{2019ApJ...870...74S}, we can just map the $\cos i$ - $M$ space, as done by \cite{2017MNRAS.465.1711B} and \cite{2019ApJ...881..165Z}. The previous discussion also implies that, within this restricted parameter space, we can safely assume that GR correctly accounts for all relativistic effects; for this reason we used the DDGR orbital model to do the mapping. We refer the reader to \cite{2017MNRAS.465.1711B} for a detailed description of how the 2-D probability density functions (pdfs) are derived.

For the \psr\, system, the 2-D pdfs are depicted in the main panels of Fig.~\ref{figure:mass_mass} by the closed black contours; these include 68.3 and 95.4 \% of the total 2-D probability, which is equivalent to the 1 and 2-$\sigma$ percentiles. The 1-D marginalisation of these 2-D pdfs along the relevant axes are presented in the side panels of that figure.

Projecting this 2-D pdf along different axes results in the mass and inclination estimates reported in the last column of Table~\ref{tab:binary}, which we also list in the abstract. These are fully consistent with the DDGR estimates, although slightly less precise.

\section{Geometry of \psr\ from pulse structure data}
\label{sec:rvm}

The variation of the position angle, $\psi$, of the linearly polarised component as shown in the lower panel of Fig.~\ref{figure:prof} is often described by the Rotating Vector Model (RVM; \citealt{Radhakrishnan&Cooke}). The RVM describes $\psi$ as a function of the pulse phase $\phi$, depending on the magnetic inclination angle, $\alpha$ and the viewing angle, $\zeta$, which is the angle between the line of sight vector and the pulsar's spin. We show its modified form as presented in \citet{jk19}:
\begin{equation}
\label{eqn:rvm}
{\rm \psi} = {\rm \psi}_{0} +
{\rm arctan} \left( \frac{{\rm sin}\alpha
\, {\rm sin}(\phi - \phi_0 - \Delta)}{{\rm sin}\zeta
\, {\rm cos}\alpha - {\rm cos}\zeta
\, {\rm sin}\alpha \, {\rm cos}(\phi - \phi_0 - \Delta)} \right)
\end{equation}
where the position angle $\psi$ increases \emph{clockwise} on the sky.  This definition of $\psi$ is opposite to the astronomical convention (also known as the ``observers'' convention or the PSR/IEEE convention defined in \citealt{psrchive_ieee}) that $\Psi$ increases counterclockwise on the sky, from North to East (cf.~\citealt{DamourTaylor1992, EverettAndWeisberg2001}).  Therefore, when fitting Eqn.~\ref{eqn:rvm} to position angles measured using the astronomical convention, as in Fig,~\ref{figure:prof}, we invert the sign of the numerator in Eqn.~\ref{eqn:rvm}.  See \cite{ksv+21} for details.

In the above, $\phi_0$ is the pulse longitude at which $\psi=\psi_{0}$ and $\zeta=\alpha+\beta$, where $\beta$ is the minimum impact angle of the line of sight with respect to the magnetic axis. The additional term $\Delta$ is is defined as 

\[
 \Delta(\phi)= 
\begin{cases}
    \Gamma,         & a \leq \phi \geq b\\
    0,              & \text{otherwise}
\end{cases} 
\]
where $a$ and $b$ are the start and end of the range of phases identified as the pre/post-cursor and $\Gamma$ is the free parameter that allows a shift in longitude (owing to a variation of either the emission height or the refractive properties of the magnetospheric plasma)

Performing a fit of Eqn.~\ref{eqn:rvm} using the method of \cite{jk19} to the position angle of \psr{} as shown in Fig.~\ref{figure:prof}, we obtain for $\Delta \equiv 0$, $\alpha = 73.7 \pm0.6$ deg and $\zeta = 77.4\pm 0.7$ deg. Interestingly, the determined $\phi_0=86.5\pm0.1$ places the fiducial plane (given by the magnetic and rotation axes and the line-of-sight to the observer) at a longitude where the circular polarisation shows its sense reversal, giving the geometrical interpretation of the position angle already some credibility. In the resulting solution, the pre/post-cursor's position angle is separated from the main pulse by an orthogonal shift of 90 deg, which is not uncommon for emission from the opposite pole (e.g.~\citealt{jk19}). 

For binaries where the spin of the pulsar is aligned with the orbital angular momentum, in the above definition of angles, $\zeta \equiv i$ (i.e.~$i$ is defined as implemented in {\sc Tempo} or {\sc Tempo2}, see \citealt{ksv+21} for details). Our measurement of the inclination angle from timing (see Table~\ref{tab:binary}) of $i=83.2\pm 0.4$ deg is significantly different from the value of $\zeta$ obtained from RVM fits. Allowing $\Gamma$ to obtain a non-zero value in the fit, i.e.~separating the emission heights of main pulse and the pre/post-cursor, does not improve the fit (as indicated by computing the Bayesian Information Criterion) and yields a very similar geometry, $\alpha = 72.4 \pm0.9$ deg and $\zeta = 76.2\pm 0.9$. The resulting $\Gamma = -6 \pm 4$ deg is still consistent with no shift, so that we will assume $\Delta\equiv0$ in the following.

\begin{figure*}
 \centering
 \includegraphics[width=\textwidth]{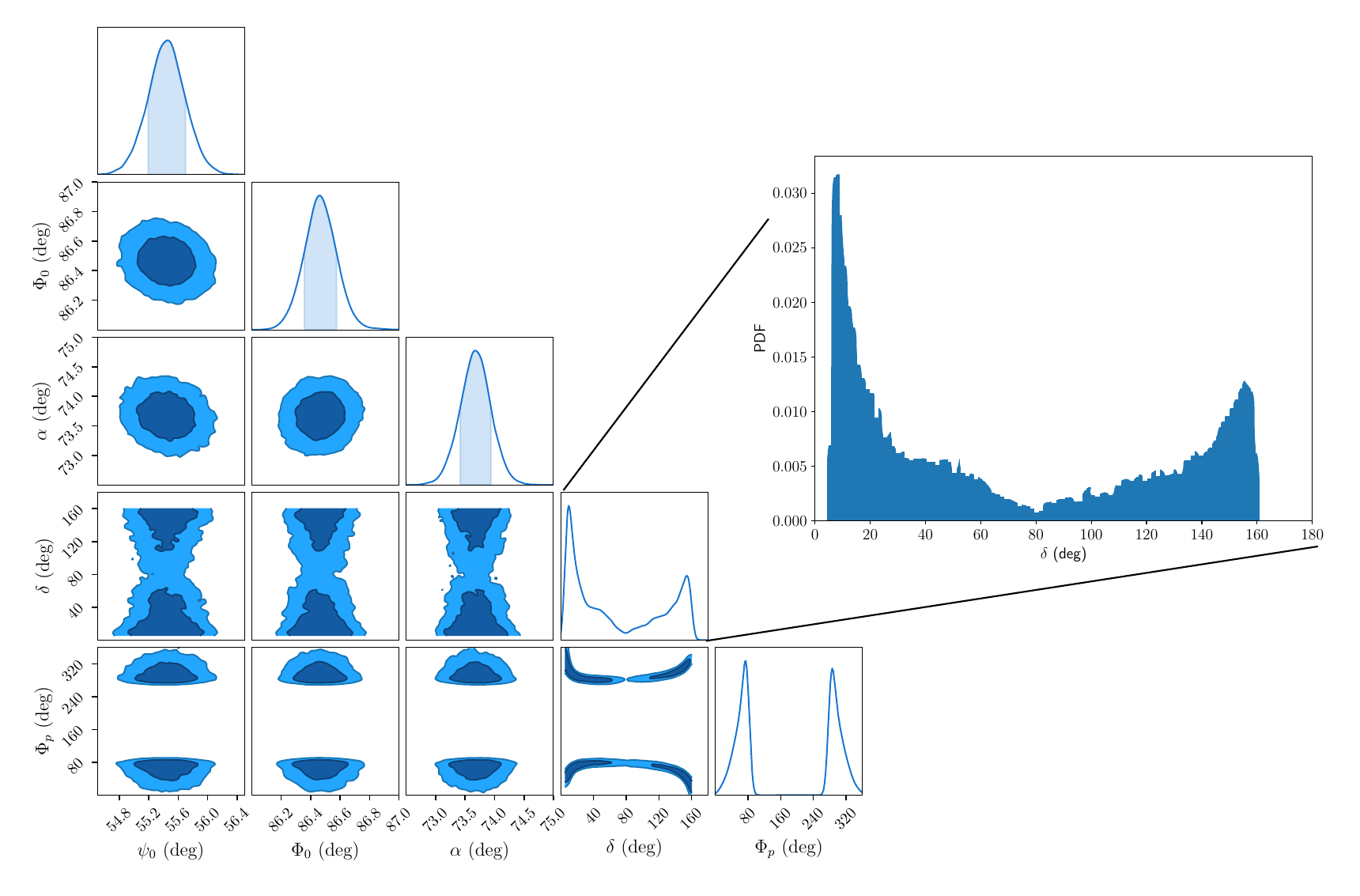}
 \caption{Posterior distributions resulting from applying the Rotating Vector Model (RVM) to the observed position angle variation, allowing for a non-zero misalignment between the spin axis and the orbital momentum vector, ie, $0< \delta < 180$ deg.}
 \label{fig:rvmposteriors}
\end{figure*}

It is of course possible that the position angle swing of recycled pulsars is not well described by a RVM. The clear deviation of the measured position angles in the longitude range at about 100 deg may be an indication of this. Indeed, there are clearly a number of average pulse profiles of recycled pulsars that are apparently difficult, or impossible, to describe with a RVM. We refer to the recent discussion in \cite{ksv+21} for more details. In such cases, it may be possible that the underlying magnetic field geometry may be non-dipolar (e.g.~caused by sweep-back of the magnetic field lines in the compact magnetosphere of millisecond pulsars that lead naturally to large emission emission heights relative to the light cylinder). Alternatively, an average profile may mask underlying orthogonal jumps in the position angle that distort the measured average, as known from non-recycled pulsars (e.g.~\citealt{gl95}). Interestingly, the deviation from the RVM fit around longitude 100 deg is indeed where the model places a transition between two orthogonal branches of the RVM.  We note that the emission over a large longitude range, especially the position angles of the nearly completely polarised pre/post-cursor provide a significant ``leverage arm'' that is able to constrain possible geometries very significantly, as it is well known from interpulse pulsars (e.g.~\citealt{jk19}). Moreover, recently, despite the overall difficulty in describing the position angles of recycled pulsars with RVMs,  a number of cases has been presented where the orbital inclination angle determined from RVM-fits was indeed in very good agreement with the value inferred from pulsar timing (see e.g.~\citealt{gfg+21} and \citealt{ksv+21}). Motivated by these previous findings, and with the described caveats in mind, we investigate if the difference between $\zeta$ and $i$ could also be due to our assumption of spin-orbit alignment. Given the unknown nature of the binary evolution of eMSPs, it is prudent to consider the possibility that the spins are indeed not aligned. In such a case, we can relate $\zeta$ and $i$ more generally following \cite{DamourTaylor1992}: 
\begin{equation} \label{eqn:sg1}
\cos\zeta = \sin \delta  \sin(180- i) \cos \Phi_p - \cos (180-i) \cos \delta ,
\end{equation}
where $\delta$ is the misalignment angle between the pulsar spin axis and the orbital momentum vector.\footnote{Note that in \cite{DamourTaylor1992}, the inclination angle is defined as $i_{\rm DT92}=180-i$, which we accounted for.} In the case of $\delta>0$, the pulsar spin will precess about the orbital angular momentum vector with a phase angle $\Phi_p$, as it has been observed for a number of relativistic double neutron star systems (see e.g.~\citealt{kra14}). Hence,
\begin{equation} \label{eqn:sg3}
 \Phi_p = \Phi_0 +  \Omega_{\rm geod} ( t-t_0),
\end{equation} 
In our case of \psr, the expected precession rate, $\Omega_{\rm geod}$, will be negligible, and over the timing baseline of $\sim 10$ years we can safely assume $\Phi_p= \Phi_0 = const$. With this in place we perform Markov-Chain Monte-Carlo (MCMC) fits to the position angle curve applying Eqns. \ref{eqn:rvm} and \ref{eqn:sg1} simultaneously. We assume $i$ to be identical to the value determined by pulsar timing (or $180-i$ deg, respectively). The posterior distributions of the model parameters are shown in Fig.~\ref{fig:rvmposteriors}. It is interesting to note that the posterior distribution of $\delta$, while broad, has two clear peaks just below 10 deg and around 160 deg. In both (prograde and retrograde) cases, $\delta$ differs significantly from $\delta=0$ or 180 deg, respectively. The first peak has a much larger amplitude. Using its location as our most likely value, we obtain $\delta = 6.1$ deg while if we constrain for the prograde case ($0\le\delta\le 90$ deg), we obtain $\delta > 4.8$ deg with 99\% CI. This result suggests that there could indeed be (prograde) spin-orbit misalignment in this system. We note that when inspecting the retrograde solution ($90\le\delta\le180$ deg), we find the peak of the PDF as $\delta = 159.3$ deg and $\delta < 160.8$ deg with 99\% CI. The ratio between the peak probability of $\delta$ for the prograde and retrograde case is 1.59.

\begin{table*}[h]
 \label{table:rvm}
 \caption[]{Measured parameter values and their 68.3\% C. L. for the different kinds of Rotating Vector Model (RVM) fits to the position angle of the linear polarisation of \psr, as shown in Fig. \ref{figure:prof}.}
\def\arraystretch{1.2}
 \centering
  \begin{tabular}{l c c c c c c r}
   \hline
   \hline
   Description & Magnetic & Co-latitude  & Impact angle& Emission & Spin misalignment &\\
   & inclination,& of spin axis, &  of line of sight, & height offset,& angle, & Technique\\   
   & $\alpha$ (deg) &$\zeta$ (deg)   &$\beta$ (deg)  & $\Gamma$ (deg) &$\delta$ (deg) & Reference \\ 
   \hline
   Classic RVM &$73.7 \pm 0.6$&$77.4 \pm 0.7$&$3.7 \pm 0.9$&$\equiv 0$&$\equiv 0$& 1 \\
   RVM with varying&$72.4 \pm 0.9$&$76.2 \pm 0.9$&$3.8 \pm 1.3$&$-6 \pm 4$&$\equiv 0$& 2\\
   emission height  &&&&&& \\
   RVM with spin-  & $73.7 \pm 0.3$ & $77.4\pm 0.3$ & $3.7\pm 0.5$ &$\equiv 0$& $6.1^{+16}_{-1.3}$$\dagger$ & 3\\
   orbit misalignment  &&&&&&\\ 
   \hline 
  \end{tabular}
  \tablebib{(1)~\cite{Radhakrishnan&Cooke}, (2)~\cite{jk19}, (3)~This work; See Section \ref{sec:rvm}}
  \tablefoot{$\dagger$ This value assumes a likely prograde misalignment. The slope of the distribution to the left of the maximumum likelihood point is so steep
  that after integrating about 6\% of the probabilities (until 4.8 degrees), the distribution ends. Since in all our measurements, we quote 68\% confidence interval by integrating 34\% of the probabilities on each side of the maximum likelihood point, we can consider 4.8 degrees as both the 68\% and 99\% lower limit. }
\end{table*}

In order to verify our method, we repeated the exact same procedure for PSR J1811$-$2405 whose original RVM fits were presented in \cite{ksv+21}. Following a similar approach, we obtain a distribution of $\delta$ that is well in agreement of spin-orbit alignment, i.e.~the corresponding distribution peaks at $\delta=0$ as expected from circular He-WD binaries.

\section{Discussion} \label{section:Discussion}

Looking at Table~\ref{table:eccentric_MSPs}, we see that the mass of \psr\, is intermediate between that of PSR~J1950+2414 ($\sim 1.50\,M_{\odot}$) and that of PSR~J1946+3417 ($\sim 1.83\,M_{\odot}$). The NS mass distribution observed in these systems seems to be similar to that observed among MSPs in systems with He~WDs in general. This is what should be expected according to both \cite{2014ApJ...797L..24A} and \cite{2021ApJ...909..161H}. This reinforces the idea that eMSPs have a broad range of masses, which disfavours the hypotheses associated with sudden phase transitions in the interior of the MSP or its super-Chandrasekhar WD progenitor as the origin of the enhanced eccentricity. As mentioned previously, e.g. the RD-AIC hypothesis predicts that all MSPs formed that way should have a mass smaller than $\sim 1.3 \, M_{\odot}$, which is not observed. The internal phase transition theory does predict  larger NS masses, but with a relatively narrow range, which is not observed either.

However, as we will see below, our new measurement ($0.254(2)\,M_\odot$) of the mass of the WD companion to \psr\, independently disfavours {\em all} hypotheses suggested to date for the formation of eMSPs. The reason is that its mass is significantly smaller than the predictions of \cite{1999A&A...350..928T} for its orbital period, and this $P_{\rm b}-M_{\rm WD}$ correlation is the backbone in all hypotheses for the formation of eMSPs. More specifically, using eqs.~(20+21) in \cite{1999A&A...350..928T}, for an orbital period of 24.58~d, the predictions for $M_{\rm c}$ vary between $\sim 0.271\, M_{\odot}$ for Population~I progenitors (corresponding to a metallicity of $Z=0.02$ and independent of the initial mass of the low-mass progenitor star) and $\sim 0.300\, M_{\odot}$ for a Population~II progenitor ($Z=0.001$).

At first sight, a WD mass deviation of order $0.02\;M_\odot$ may not seem like a lot. However, the difference of $0.017\;M_{\odot}$ relative to the lower limit of the $P_{\rm b}-M_{\rm WD}$ predictions, is 7.4 times larger than the measurement uncertainty. And more importantly, for the observed $M_{\rm c}$, the corresponding orbital periods are only about 14~d for Population~I progenitors of the He~WD and 6~d for Population~II progenitors. That is a very significant deviation from the observed value of 24.58~d. Previous comparison with wide-orbit binary MSPs, although for binaries without precise mass measurements, has indicated that the $P_{\rm b}-M_{\rm WD}$ correlation may overestimate the WD masses \citep{StairsEtAl2005}. The small WD mass of \psr\, may possibly be explained by an unusual high metallicity content of its progenitor star or due to incorrect input physics in current modelling of the correlation. However, it is important to mention that independent theoretical support for the applied $P_{\rm b}-M_{\rm WD}$ correlation has been provided by several more recent studies \citep[e.g.][]{2011ApJ...732...70L,2016A&A...595A..35I,sk21}; see Fig.~14.14 in \citet{tv22} for an updated compilation of data.

The new result for \psr\, is unique among eMSPs, for which previously measured He~WD masses conform to the relation found by \cite{1999A&A...350..928T} --- although the companion to PSR~J1946+3417 is also slightly less massive than the prediction. The small value of $M_{\rm c}$ for, in particular, \psr\, is a highly significant result, and yet puzzling as all the suggested hypotheses naturally produce the observed range of orbital periods of eMSPs. We will now look at possible explanations for this, and its implications.

\subsection{On the measurement precision and reliability of $\dot{\omega}$}

A possibility is that our measurement of $\dot{\omega}$ is hampered by the low quality of earlier Parkes measurements. If this were true, then future MeerKAT measurements are bound to correct this situation very quickly. Furthermore, when we exclude the earliest filterbank data (including only the more reliable coherently dedispersed data taken since June 2015), we obtain very similar masses to those reported above: $M_{\rm c} \, = \, 0.253(3)$ and $M_{\rm p} \, = \, 1.70(3)\, M_{\odot}$. Including only the high-quality MeerKAT and Parkes data taken since 2019, we obtain $M_{\rm c} \, = \, 0.252(3)$ and $M_{\rm p} \, = \, 1.69(3)\, M_{\odot}$. All these values are consistent well within their uncertainties; this means that those early data do not have a significant weight on our mass estimates, and certainly they do not bias them appreciably.

\subsection{Mass loss from the WD?}

Since our mass measurement appears to be reliable, the deviation from the predictions of \cite{1999A&A...350..928T} is real. A possible explanation is that the mass deficiency of the companions is caused by ablation by the pulsar winds. Indeed, the two systems with larger mass deficiencies, \psr\,  and J1946+3417, are associated with $\gamma$-ray emission (see \citealt{CamiloEtAl2015,2019ApJ...871...78S}). Again, this is difficult to reconcile with the fact that no significant outgassing and ablation is seen in any of the eccentric MSP+He~WD systems. Such outgassing would be readily detected by the existence of eclipses and DM variations, such as those observed in many eclipsing binary pulsars. Such phenomena are not detectable in our observations of \psr\, at any orbital phase, even at superior conjunction.

\subsection{Something unique to eccentric MSPs?}

If mass loss from the WD cannot explain the measured low mass, we must entertain the possibility that this low mass is somehow linked to the fact that this system is an eccentric MSP. First, given the ability to measure $\dot{\omega}$ in these systems, we can obtain unusually precise measurements of the masses of their pulsars and He~WDs. Thus, it is in principle possible that the existence of under-massive He~WDs is a common occurrence among the circular MSP+He~WD systems; a fact that could have been undetected until now because of the low precision for the measurements of the masses in the vast majority of those systems. However, this is unlikely, since there is strong independent evidence that the T\&S99 relation really is universal (see e.g., \citealt{2014ApJ...781L..13T}): even the highly precise measurements for the two WDs in the triple system \citep{2020A&A...638A..24V} conform exactly to the prediction of \cite{1999A&A...350..928T}.

Thus we conclude the low mass of the companion to \psr\,  could be an important, but thus far hard to interpret, clue on the poorly understood formation of eMSPs. However, as discussed below, this measurement rules out current hypotheses for the formation of these systems, if one assumes that the unique orbital period--mass correlation for He~WDs \citep{1987Natur.325..416S,1995MNRAS.273..731R,1999A&A...350..928T,2011ApJ...732...70L,2016A&A...595A..35I}, which is based on the well-known correlation between stellar radius and degenerate core mass for low-mass giant stars \citep{1971A&A....13..367R}, holds during the formation of MSP+He~WD systems.

\begin{figure*}
 \centering
 \includegraphics[width=0.60\textwidth, angle=-90]{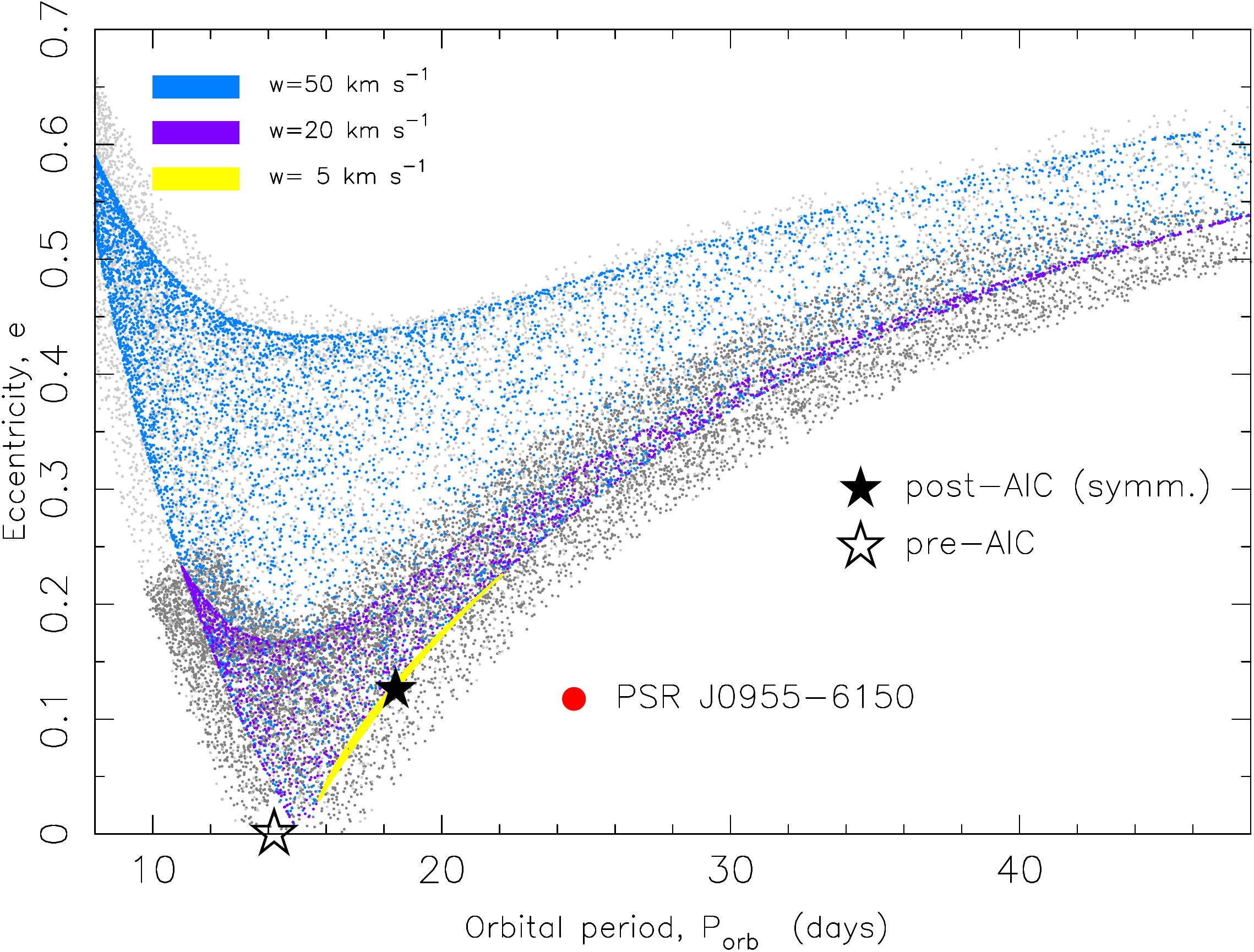}
 \caption{Distribution in the $(P_{\rm b},\,e)$--plane of eMSPs simulated via the RD-AIC scenario. The red circle represents \psr. The black solid star is the result of a symmetric ($w=0$) AIC from a $1.96\;M_{\odot}$ super-Chandrasekhar WD with a pre-AIC orbital period, $P_{b,0}=14.2\;{\rm d}$ (open star) and a He~WD companion star of mass, $M_{\rm c}=0.254\;M_{\odot}$. The V-shaped coloured distributions are for the same system but applying a kick of $w=5$, 20, and $50\;{\rm km}\,{\rm s}^{-1}$, respectively, in a random (isotropic) direction. The grey and light-grey distributions superimposed are the outcome of simulations with more relaxed assumptions on the input physics parameters --- see text. The RD-AIC scenario fails to explain \psr.}
 \label{figure:RD+AIC}
\end{figure*}

\subsection{\psr\, independently rules out phase-transition models}

According to the $P_{\rm b}-M_{\rm WD}$ correlation from \citep{1999A&A...350..928T}, the companion star mass of {$M_{\rm c}\simeq 0.254\;M_\odot$ for \psr\, reveals that this He~WD companion must have had an orbital period between $P_{b,0}=6.19\;{\rm d}$ (Pop.~II) and $P_{b,0}=14.2\;{\rm d}$ (Pop.~I)} at the time of its formation, depending on the chemical composition of its progenitor star \citep[see eqs.~20+21 in][]{1999A&A...350..928T}.

Following as an example the RD-AIC scenario, from the current pulsar mass of $M_{\rm p}\simeq 1.71\;M_\odot$, we can estimate the mass of its super-Chandrasekhar mass WD progenitor, taking into account the loss of gravitational binding energy in the AIC process. The NS binding energy depends on the still somewhat uncertain equation-of-state. Here, we adopt the binding energy calculation from \citet{1989ApJ...340..426L}, which as a result reveals a pre-AIC WD mass, $M_{\rm AIC}=1.96\;M_\odot$, i.e. corresponding an instantaneous mass loss in the AIC process of about $\Delta M \simeq 0.246\;M_\odot$.

Given now the pre-AIC orbital period (depending on metallicity) and mass of the collapsing WD, we can calculate the post-AIC orbital parameters and compare to those of \psr. The post-AIC eccentricity (assuming no kick at birth) is simply given by: $e=\Delta M/M_{\rm T}$ (where $M_{\rm T}=M_{\rm p}+M_{\rm c}$), which for the derived masses yields $e\simeq 0.125$. This value is only marginally larger than the observed eccentricity of $e=0.1175$; and adopting a slightly smaller value for the binding energy ($\Delta M=0.231\;M_\odot$) would easily reproduce the exact value of the observed eccentricity. However, the problem is to reproduce the post-AIC (present) orbital period. Assuming first a high metallicity environment ($Z=0.02$, i.e. Pop.~I), means that the pre-AIC orbital separation was $P_{b,0}=14.2\;{\rm d}$ from the orbital period--mass correlation. For a symmetric collapse (i.e. with no kick), the change in orbital period is given by\footnote{See e.g. \cite{tv22}.}:
\begin{equation}
\frac{P_{\rm b}}{P_{b,0}} = M_{\rm T}\,\sqrt{\displaystyle\frac{M_{\rm T}+\Delta M}{(M_{\rm T}-\Delta M)^3}} 
\label{eq:P_SN-symm}
\end{equation}
such that the present (post-AIC) orbital period should be $P_{\rm b}=18.4\;{\rm d}$, which is significantly smaller than the observed value of 24.58~d. This discrepancy is only exacerbated if we assume a low-metallicity (Pop.~II) chemical abundance of the progenitor star of the current He~WD, which would then produce a post-AIC orbital period of only 8.05~d. Note that no orbital evolution of \psr\, has taken place since its formation. The tidal torques and rate of circularization due to GR are minuscule.

To explore whether an applied momentum kick to the newborn NS could resolve the problem of reproducing the observed values of ($P_{\rm b},\,e$), we ran a number of numerical Monte Carlo simulations. We assumed the pre-AIC system was circular with an orbital period of $P_{b,0}=14.2\;{\rm d}$ (Pop.~I star progenitor). 

Figure~\ref{figure:RD+AIC} shows the results of our simulations. The open star marks the pre-AIC system. The solid star marks the post-AIC system for a symmetric ($w=0$) AIC event. The yellow, purple and light blue points are the outcome of AIC events with a kick velocity of $w=5$, 20 and $50\;{\rm km\,s}^{-1}$, respectively, and assuming a random (isotropic) direction of the kick. The equations governing the outcome are found in e.g. \citet{hil83}. We see that none of the simulated systems come close to the parameter space in vicinity of \psr, and thus these simulations based on the RD-AIC model are not successful in explaining the formation of this eMSP.

To take into account the uncertainties in the gravitational binding energy for the NS, and also for the orbital period--mass correlation, we ran two extra sets of simulations, where, in both cases: the pre-AIC mass ($M_{\rm AIC}$) was randomly drawn from a flat probability distribution between $1.86-2.06\;M_\odot$; the pre-orbital period ($P_{b,0}$) was randomly drawn from a flat probability distribution between $12.2-16.2\;{\rm d}$; and the mass of the present He~WD ($M_{\rm c}$) was randomly drawn from a flat probability distribution between $0.251-0.257\;M_\odot$. The grey and light grey points show the resultant post-AIC systems assuming randomly-directed kicks of $w=20\;{\rm km\,s}^{-1}$ and $w=50\;{\rm km\,s}^{-1}$, respectively, during the AIC events. Even choosing such large AIC kicks is probably unrealistic \citep[see e.g.][]{2006ApJ...644.1063D,2018ApJ...865...61G}. Nevertheless, even relaxing generously on the assumed physical parameters prevents us from reproducing a system similar to \psr. 

Therefore, we conclude that the RD-AIC hypothesis of \citet{2014MNRAS.438L..86F} can no longer be considered a potential model for explaining the existence of eMSPs. Whereas the RD-AIC model could explain very well the formation of the eMSPs known at that time, PSR~J2234+0611 and PSR~J1946+3417, it fails to explain \psr\, due to its relatively low-mass He~WD companion, which dictates a short orbital period that cannot be reproduced.
The same arguments can be used to rule out the internal phase transition model proposed by \cite{2015ApJ...807...41J}, since it predicts similar constant losses in binding energy during the phase transition,

\subsection{\psr\, rules out H-shell flash models}\label{subsec:rule-out-flashes}

According to \cite{2014ApJ...797L..24A}, the eMSPs were produced like regular MSP+He~WD systems, which likewise follow the predictions of \cite{1999A&A...350..928T}. Under this hypothesis, the orbital eccentricity was caused by eccentricity pumping via a circumbinary disk of material ejected by H-shell flashes in the outer layers of the proto-He~WD. These H-shell flashes are likely to happen in a wide range of He~WD masses between $\sim \, 0.16-0.32\;M_{\odot}$ \citep[depending on metallicity,][]{2013A&A...557A..19A,2016A&A...595A..35I}, which should then occur in the corresponding range of orbital periods observed for the eMSPs.

The expected mass loss via RLO from such vigorous thermonuclear runaway episodes is only of the order of $10^{-5}$ to $10^{-3} \, M_{\odot}$, a difference that cannot, by itself, explain the small mass of the companion to \psr. Furthermore, the ejection of such a small amount of matter is unlikely to significantly change the semi-major axis of the binary. Depending on the specific orbital angular momentum carried away by the ejected material, if anything, this ejection of should actually {\em decrease} the orbit, thus bringing the new orbital period closer to the \cite{1999A&A...350..928T} prediction for the slightly decreased He~WD mass. Thus, even after acquiring the new eccentricity, the system should retain values of $M_{\rm c}$ and $P_{\rm b}$ close to the relation predicted by \cite{1999A&A...350..928T}.

Regarding the remaining two recent hypotheses by \citet{2021ApJ...909..161H}, related to H-shell flashes (thermonuclear rocket effect), and \citet{Ginzburg21} on resonant convection, it is not obvious that they are successful in the end. The former model has the advantage of begin able to explain a broad range of eMSPs (roughly $18-45\;{\rm d}$ according to the authors), but the assumption of an instantaneous kick may not apply in reality. Furthermore, the predicted range of WD masses for which H-shell flashes are expected \citep{2016A&A...595A..35I} goes much beyond the narrow range of WD masses for eMSPs. The latter model has perhaps the weakness of not explaining well why some MSP binaries with similar orbital periods as the eMSP did not experience resonant interactions and remained in circular orbits with $e\simeq 10^{-5}$ \citep[see fig.~1 in][]{2019ApJ...870...74S}. Nor is it clear if the model can explain the offset of \psr\, from the $P_{\rm b}-M_{\rm WD}$ correlation.

\subsection{Is the misalignment angle a clue to the origin of \psr?}\label{subsec:clue}

It is expected from binary star evolution that the spin axis of the MSP aligns with the orbital angular momentum vector as a result of mass transfer. Here we follow the arguments by \citet{tv22}. During RLO, accretion torques align the spin axis of the first-born compact object (here the NS) with the orbital angular momentum vector during the recycling process \citep[e.g.][]{hil83,bv91,ba21}. Observational evidence for such an alignment to actually occur in nature was demonstrated for LMXBs by \citet{gt14}, who found agreement between the viewing angles of binary MSPs (as inferred from $\gamma$-ray light-curve modelling) and their orbital inclination angles. Therefore, it is reasonable to assume $\delta\simeq 0$ for recycled MSPs (i.e. that the MSP spin axis is, at least close to, parallel to the orbital angular momentum vector of the binary system.

In this work, we have demonstrated that for \psr\, if assuming an ideal dipolar magnetic field, then, surprisingly, $\delta > 4.8 \deg$. The combined unexpected result of a significant misalignment angle and this MSP being an eMSP, makes it tempting to suggest that there might be a relation between between these two circumstances. To test this idea, it is therefore important to measure (or significantly constrain) the value of $\delta$ for all other eMSP systems. In addition, we may ask: which progenitor scenarios may account for such misalignment? At first sight, the ``thermonuclear rocket'' hypothesis of \cite{2021ApJ...909..161H} may be a natural way to explain such a misalignment, if indeed the orbit is tilted as a result of thermonuclear runaway burning events (e.g. H-shell flashes).

To investigate this question, we simulated a large population of NS+ELM~He~WD systems similar to \psr\, undergoing a large H-shell flash with a relatively large kick of $w=8\;{\rm km\,s}^{-1}$ (in a random direction), a large amount of ejected material of $\Delta M=10^{-3}\;M_\odot$ and a pre-shell orbital period of $P_{\rm b}=20.0\;{\rm d}$. All systems had $M_{\rm p}=1.71\;M_\odot$ and $M_{\rm c}+\Delta M=0.255\;M_\odot$ (to leave a final $0.254\;M_\odot$ WD). The resulting misalignment angles are always $\delta < 4.7^\circ$. 

Even though we adopted rather large values of $w$ and $\Delta M$ in our simulation, the ``thermonuclear rocket'' scenario of \citet{2021ApJ...909..161H} seems to produce too small misalignment angles to explain the observed value of $\delta$ in \psr. Potentially more serious is that the assumption in this hypothesis of an instantaneous kick (compared to the timescale of $P_{\rm b}$) is likely not justified. Assuming instead mass loss in the form of a fast wind (i.e. the Jeans' mode over a timescale of several times $P_{\rm b}$) would cause an orbital widening to only $\Delta P_{\rm b}\simeq 0.016\;{\rm d}$, i.e. the orbit remains more or less constant with a negligible eccentricity increase.  
Finally, as discussed in Section~\ref{subsec:rule-out-flashes} and similarly to the other models discussed in this paper, there are issues with reproducing the observed orbital period for the observed WD mass in \psr\,.

We conclude therefore, that none of the formation hypotheses suggested in the literature to date is able to explain well the low mass of the He~WD companion to \psr\, and its orbital misalignment. This means that the formation of eMSPs remains a major puzzle of close binary stellar evolution.

\section{Summary and conclusions}
\label{sec:summary}

In this paper, we described our observations of \psr\, with the MeerKAT and Parkes telescopes. Previously, this pulsar did not have a phase-coherent timing solution, we report it here for the first time. The high S/N ratio and the resultant high-precision timing obtained with MeerKAT was instrumental for the detection and precise measurement of three PK parameters; the measurement of the rate of advance of periastron was also made possible by the large timing baseline and the fact that the system is unusually eccentric ($e \, = \, 0.11$) for an MSP, which makes it a member of a growing class of MSPs with He~WD companions with eccentric orbits.

The measurements of these three PK parameters yielded precise mass measurements, which we discuss in light of the different hypotheses that have been advanced to explain the formation of these unusual and intriguing systems. One intriguing finding is that the He~WD mass is significantly lower than predicted by current stellar evolution models. The significance of this finding is very important and could be a clue to new progress on the poorly understood formation mechanism for these eMSP systems. It is clear, however, that it independently rules out all of the proposed formation hypotheses presented until now, since in all these models the systems should obey the $P_{\rm b}-M_{\rm WD}$ correlation \citep{1999A&A...350..928T}.

Additionally, if we assume the pulsar's radio emission is purely dipolar where the position angle of the linear polarisation ideally follows the RVM, we find a $\delta > 4.8 \deg$ misalignment between the spin axis of the pulsar and the orbital angular momentum. While this result must be take with a pinch of salt due to the yet not fully understood nature of millisecond pulsar polarisation, it is nevertheless an interesting hint that could be solidified by analysing other eMSPs. None of the formation hypotheses predict such a large misalignment, except for the ``thermonuclear rocket'' of \cite{2021ApJ...909..161H}. Assuming the assumptions behind this hypothesis to hold true, our simple toy simulation of this model indicates that the resulting misalignment angles are likely to be too small to explain the observed value of $\delta$, although more work is needed to fully exploit this scenario. We believe that our measurement of a misalignment angle is an important datum for understanding the origin of eMSPs.

Continued observations with MeerKAT, in particular a second orbital campaign done with the UHF band, will greatly improve the precision and accuracy of the measurement of $\dot{\omega}$ and of the Shapiro delay as well, and allow a much improved determination of the masses of the components in this system and an improved $\dot{\omega}$ - $h_3$ - $\varsigma$ test of GR. Continued timing will also allow a much improved measurement of $\dot{P}_b$, this will eventually result in a much more precise distance to the system.

The polarimetry of the system at UHF bands, where the pulsar is much brighter, and especially a detailed study of the $\gamma$-ray emission of the system (Paper~II, in prep.) have the potential to further refine our knowledge of the spin geometry of the pulsar, and confirm (or refute) our finding of the misalignment of the pulsar spin with the orbital angular momentum, which relies on the validity of the RVM model for \psr. Confirming this finding, preferentially in a way that is independent of the RVM, would provide a very important and unexpected datum for understanding the origin of eMSPs.

\begin{acknowledgements}
The authors would like to thank Rosie Chen for help with analysis relating to search of the counterpart in optical and near-infrared survey data and John Antoniadis for the discussions on the evolution of the He~WD companion to \psr. The MeerKAT telescope is operated by the South African Radio Astronomy Observatory, which is a facility of the National Research Foundation, an agency of the Department of Science and Innovation. SARAO acknowledges the ongoing advice and calibration of GPS systems by the National Metrology Institute of South Africa (NMISA) and the time space reference systems department department of the Paris Observatory. MeerTime data is housed on the OzSTAR supercomputer at Swinburne University of Technology. The Parkes radio telescope is funded by the Commonwealth of Australia for operation as a National Facility managed by CSIRO. We acknowledge the Wiradjuri people as the traditional owners of the Observatory site. This research has made extensive use of NASAs Astrophysics Data System (https://ui.adsabs.harvard.edu/) and includes archived data obtained through the CSIRO Data Access Portal (http://data.csiro.au). Parts of this research were conducted by the Australian Research Council Centre of Excellence for Gravitational Wave Discovery (OzGrav), through project number CE170100004 and the Laureate fellowship number FL150100148. VVK, PCCF, MK, AP and MCiB acknowledge continuing valuable support from the Max-Planck Society. APo and MBu acknowledge the support from the Ministero degli Affari Esteri e della Cooperazione Internazionale - Direzione Generale per la Promozione del Sistema Paese - Progetto di Grande Rilevanza ZA18GR02. MBu and APo acknowledge support through the research grant "iPeska" (PI: Andrea Possenti) funded under the INAF national call Prin-SKA/CTA approved with the Presidential Decree 70/2016. RMS acknowledges support through Australian Research Council Future Fellowship FT190100155. Pulsar research at UBC is supported by an NSERC Discovery Grant and by the Canadian Institute for Advanced Resarch. This publication made use of open source python libraries including Numpy \citep{numpy}, Matplotlib \citep{matplotlib}, Astropy \citep{astropy} and Chain Consumer \citep{Hinton2016_ChainConsumer}, along with pulsar analysis packages: \textsc{psrchive} \citep{HotanEtAl2004}, \textsc{tempo2} \citep{HobbsEtAl2006}, \textsc{temponest} \citep{LentatiEtAl2014}.
\end{acknowledgements}

\bibliographystyle{aa}
\bibliography{mserylak_j0955.bib}

\end{document}